\documentclass[a4paper,11pt]{article}
\pdfoutput=1 
\usepackage{jcappub} 

\usepackage{subcaption}
\usepackage[T1]{fontenc} 
\usepackage{verbatim}

\usepackage[utf8]{inputenc}

\usepackage[style=nature,backend=biber, sorting = none, doi=false, maxbibnames=8, minbibnames=3]{biblatex}

\usepackage[dvipsnames]{xcolor}

\newcommand{\simgt}%
{\,\hbox{\lower0.6ex\hbox{$\sim$}\llap{\raise0.6ex\hbox{$>$}}}\,}
\newcommand{\simlt}%
{\,\hbox{\lower0.6ex\hbox{$\sim$}\llap{\raise0.6ex\hbox{$<$}}}\,}
\newcommand{\hompc}{\,h\,{\rm Mpc}^{-1}}
\newcommand{\mpcoh}{\,h^{-1}\,{\rm Mpc}}

\title{\textbf{Testing Large-Scale Structure Measurements Against Fisher Matrix Predictions}}

\author[a,b]{Setareh Foroozan}
\author[a,b,c]{Alex Krolewski}
\author[a,b,c]{Will J. Percival}

\affiliation[a]{Waterloo Centre for Astrophysics, University of Waterloo, 200 University Ave W, Waterloo, ON N2L 3G1, Canada}
\affiliation[b]{Department of Physics and Astronomy, University of Waterloo, 200 University Ave W, Waterloo, ON N2L 3G1, Canada}
\affiliation[c]{Perimeter Institute for Theoretical Physics,
31 Caroline St. North, Waterloo, ON N2L 2Y5, Canada}

\emailAdd{s2forooz@uwaterloo.ca}

\abstract{
We compare Baryonic Acoustic Oscillation (BAO) and Redshift Space Distortion (RSD) measurements from recent galaxy surveys with their Fisher matrix based predictions. Measurements of the position of the BAO signal lead to constraints on the comoving angular diameter distance $D_{M}$ and the Hubble distance $D_{H}$ that agree well with their Fisher matrix based expectations. However, RSD-based measurements of the growth rate $f \sigma_{8}$ do not agree with the predictions made before the surveys were undertaken, even when repeating those predictions using the actual survey parameters. We show that this is due to a combination of effects including degeneracies with the geometric parameters $D_{M}$ and $D_{H}$, and optimistic assumptions about the scale to which the linear signal can be extracted. We show that measurements using current data and large-scale modelling techniques extract an equivalent amount of signal to that in the linear regime for $k\simlt 0.08 \hompc$, remarkably independent of the sample properties and redshifts covered.}

\begin{document}
\maketitle
\flushbottom

\section{Introduction}
\label{sec:intro}

The quest to understand Dark Energy, the physical mechanism behind observations of the accelerating expansion of the Universe, has led to a plethora of ongoing and future experiments, including the Dark Energy Spectroscopic Instrument (DESI, \cite{desi_collaboration_desi_2016}), the Rubin Observatory and LSST survey \cite{lsst_dark_energy_science_collaboration_large_2012}, and the Euclid \cite{euclid_collaboration_euclid_2020} and WFIRST \cite{WFIRST-2015-report} satellite missions. Many of these are designed to use the Baryon Acoustic Oscillation (BAO), and Redshift-Space Distortion (RSD) signals within the clustering of galaxies to constrain the geometry of the Universe and growth of structure within it. 

Over the past two decades, since the early signs of the baryon acoustic oscillations were seen in the 2-degree Field Galaxy Redshift Survey (2dFGRS, \cite{percival_2df_2001}) and the BAO signal was refined using the SDSS-II Luminous Red Galaxy (LRG) sample \cite{eisenstein_detection_2005} and the final release of data from the 2dFGRS \cite{cole_2df_2005}, ground based surveys have been undertaken to make BAO and RSD measurements to ever higher precision. The combination of 2dFGRS and the final SDSS-II LRG data reached a detection threshold of 3.6$\sigma$ \cite{percival_baryon_2010}, which was rapidly overtaken by early data from the Baryon Oscillation Spectroscopic Survey (BOSS, \cite{anderson_clustering_2012}), which breached the 5$\sigma$ detection threshold. Since then, the BAO technique has become one of the pillars of modern cosmology, with particularly important surveys undertaken within the Sloan Digital Sky Survey (SDSS, \cite{york_sloan_2000}). At low redshift, we have the Main Galaxy Sample (MGS, \cite{strauss_spectroscopic_2002}) using data from SDSS-I\&II (\cite{abazajian_seventh_2009}), while at higher redshift we have the SDSS-III (\cite{eisenstein_sdss-iii_2011}) BOSS, \cite{dawson_baryon_2013} and the SDSS-IV (\cite{blanton_sloan_2017}) extended Baryon Oscillation Spectroscopic Survey (eBOSS, \cite{dawson_sdss-iv_2016}). In addition, complementary measurements were made by the 6-degree Field Galaxy Survey (6dFGS, \cite{jones_6df_2009}) at low redshift, and the WiggleZ Dark Energy Survey (WiggleZ, \cite{parkinson_wigglez_2012}) at high redshift. All have released measurements at various stages of survey progress.

The observed BAO and RSD signals from these surveys have been analyzed by different groups with slightly different techniques in both configuration and Fourier space. These measurements constrain the anisotropic distance scales, $D_M$ and $D_H$, the isotropic distance scale, $D_V$, and the logarithmic growth rate of structure, $f$.
The Fisher matrix formalism has allowed cosmologists to predict the constraining power of surveys on these parameters and therefore plan for the future \cite{tegmark_measuring_1997,white_forecasting_2009,font-ribera_desi_2014,zhao_extended_2016}. This formalism was first introduced to estimate the error on model parameters in any given dataset \cite{fisher_logic_1935}, by assuming that the inverse of the Fisher matrix can be interpreted as an estimation of the covariance matrix for a Gaussian likelihood. Moreover, the Cramér-Rao inequality states that the diagonal elements of the inverse of the Fisher matrix give a lower bound on the variance of any unbiased estimator of the model parameters, in other words, the best possible errors. Thus, validating the Fisher matrix predictions made for past surveys is important to test whether survey goals were met and particularly to test the optimality of the analysis techniques used to evaluate the cosmological parameters. For instance, in a recent study by Ruggeri et al. \cite{ruggeri_how_2020}, the errors in BAO survey measurements, mocks, and Fisher matrix were compared without applying reconstruction to the density field for six galaxy surveys, finding good agreement. We find similar results for the same surveys and datasets analysed in the same way, but extend this analysis to consider further data: BAO with reconstruction and RSD measurements.

The goal of our paper is to compare the constraints recovered from the BAO and RSD measurements of various surveys with the Fisher matrix predictions for the expected error bars. The inputs to the Fisher calculations match as closely as possible that of each analysis. The outline of this paper is as follows. In Section~\ref{sec:method}, we briefly describe the Fisher matrix formalism and how it is applied in this paper. We continue with descriptions of the surveys considered here in Section~\ref{sec:Data} and the numbers used throughout this paper. In Section~\ref{sec:results}, we present our Fisher code results,  and then compare them to the observations. In particular, we evaluate these surveys' performance, comparing the Fisher-predicted errors with the precision recovered from the BAO and RSD measurements. Finally, we discuss the results in Section~\ref{sec:discussion}.

\section{Methodology}\label{sec:method}

We now briefly describe how we perform the Fisher-based predictions for a given survey. In order to match the experiments, we separately predict errors for BAO and RSD measurements. Before describing the specifics of these calculations, we introduce the general Fisher matrix methodology that is not exclusive to cosmology, but rather can be used for estimating the errors of any given dataset.

Supposing that $\Vec{x}$ is a random variable with the probability distribution $f(\Vec{x};\Vec{p})$, where $\Vec{p}$ is a vector of known parameters, the Fisher information matrix (\cite{tegmark_karhunen-loeve_1997, tegmark_measuring_1997}) corresponding to this set of variables is defined as
\begin{equation}\label{eq:Fisher}
    F_{i,j} \equiv - \left< \frac{\partial ^ 2 \ln{f}}{\partial p_i \partial p_j}\right>\,.
\end{equation}

Applying this to galaxy surveys, we wish to estimate a set of cosmological parameters $\{p_1, p_2, ...\}$ using the redshift space galaxy power spectrum, $P(k,\mu)$, and the galaxy number density, $n$, in the survey's volume, $V_{\rm sur}$. Following Tegmark (1997) \cite{tegmark_measuring_1997}, if we let the data vector $\vec{x}$ be the galaxy power spectrum for a Gaussian random field, Eq.~\ref{eq:Fisher} will yield the following expression for the Fisher matrix:
\begin{equation}\label{eq:Fisher2}
    F_{ij} = \frac{V_{\rm sur}}{4 \pi^2}\int_{-1}^{1}d\mu \int_{k_{\rm min}}^{k_{\rm max}} k^2 dk   \mathcal{F}_{ij}(k, \mu)\,,
\end{equation}
where,
\begin{equation}\label{eq:Fisher3}
    \begin{aligned}
        \mathcal{F}_{ij}(k, \mu) = \frac{1}{2} 
        \left(\frac{V_{\rm eff}}{V_{\rm sur}}\right)
        \frac{\partial \ln{P}}{\partial p_i} \frac{\partial \ln{P}}{\partial p_j}\,,  \\
        V_{\rm eff} = \left[\frac{n P(k,\mu)}
        {nP(k,\mu) +1}\right]^2 V_{\rm sur}\,,
    \end{aligned}
\end{equation}
and $V_{\rm eff}$ and $\mu$ are the effective volume and the cosine of the angle between $\vec{k}$ and the line of sight. In the linear regime, the power spectrum can be written
\begin{equation}
    P = P(k,\mu) = (b+f\mu^2)^2 P_{\rm lin}(k)\,,
    \label{eq:powerspectrum}
\end{equation}
where $b$, $f$, $P_{\rm lin}$ denote the galaxy bias, logarithmic growth rate and linear power spectrum.

We split each survey in $N_z$ slices, and numerically integrate Eq.~\ref{eq:Fisher2} in each redshift slice and eventually add the Fisher matrices to yield the inverse of the total covariance matrix. Moreover, for each survey, we assume the same fiducial cosmology as quoted in their corresponding BAO and RSD measurement paper. For completeness, we list these in Table~\ref{tab:summary}. In Section~\ref{subsec:Fisher_BAO} and~\ref{subsec:Fisher_RSD} we describe the constraints recovered from BAO and RSD analyses respectively.

\subsection{The Fisher matrix for the BAO measurements}\label{subsec:Fisher_BAO}

To predict the constraints on the parameters recovered from BAO measurements, we adapt the approach described in Seo \& Eisenstein \cite{seo_improved_2007}. Following their method, we construct the Fisher matrix constraints on angular diameter distance $D_M$, and the Hubble distance $D_H$, meaning that the free parameters in Eq.~\ref{eq:Fisher2} are $\{p_1, p_2\} = \{\ln{D_M}, \ln{D_H}\}$. 
These parameters can be related to the BAO dilation parameters as follows

\begin{equation}
    \begin{aligned}
        \alpha_{\perp} = \frac{D_M(z_{\rm eff})/r_{\rm drag}}{D_M^{\rm fid}(z_{\rm eff})/r_{\rm drag}^{\rm fid}}\,,
        \\
        \alpha_{\parallel} = \frac{D_H(z_{\rm eff})/r_{\rm drag}}{D_H^{\rm fid}(z_{\rm eff})/r_{\rm drag}^{\rm fid}}\,.
    \end{aligned}
\end{equation}
From these parameters we can further obtain the isotropic volume-averaged distance, $D_V = \left[zD_H(z)D_M(z)^2 \right]^{1/3}$. 
The final expression for the Fisher matrix of the anisotropic distances, as described in Seo \& Eisenstein, is
\begin{equation} \label{eq:SE}
    F_{i,j} = V_{\rm sur} A_0^2 \int_{0}^{1} d\mu f_{i}(\mu) f_{j}(\mu) \int_0^\infty dk \, \frac{k^2\exp{\left[-2(k\Sigma_s)^{1.4}\right]}}{\left(\frac{P(k)}{P_{0.2}} + \frac{1}{nP_{0.2}R(\mu)}\right)^2} \, \exp{\left[ -k^2(1-\mu^2)\Sigma_{\perp}^2 - k^2 \mu^2 \Sigma_{\parallel}^2\right]},
\end{equation}
where $P_{0.2}$ is the galaxy power at $k = 0.2 \hompc$, $\Sigma_{\perp}$ and $\Sigma_{\parallel}$ are the rms radial displacement across and along the line of sight, $\Sigma_{s}$ is the inverse of the Silk-damping scale, and $A_0$ is the normalization of the baryonic term in the Eisenstein \& Hu power spectrum \cite{eisenstein_baryonic_1998}. Depending on which element of the Fisher matrix is being calculated, $f_1(\mu) = \mu^2 - 1$ and $f_2(\mu) = \mu^2$. They assumed redshift distortions of the form 
\begin{equation}
    R(\mu) = (1+ f \mu^2 / b)^2 \exp{(-k^2\mu^2\Sigma_z^2)} \,,
\end{equation}
where $f$ is the logarithmic derivative of the linear growth rate with respect to scale factor, $\mathrm dD(a) / d \ln{(a)}$, and $b$ is the galaxy bias. The exponential term corresponds to a Gaussian uncertainty in redshift characterized by $\Sigma_z$, which will be discussed more in Section~\ref{sec:Data}.

Since the normalisation and baryon damping terms in the power spectrum $P(k)$ in Eq.~\ref{eq:SE} are functions of $\Omega_{b}$, $\Omega_{m}$, and $h$, it is important to recalculate these for the cosmology assumed if it is different from the default in the version of the code publicly released by Seo \& Eisenstein. As we adjust our Fisher calculations to match the cosmology assumed by different authors in their analyses, we have extended the code to allow the relevant parameters to change, using the Eisenstein \& Hu (1998) fitting function for the power spectrum. Additionally, since the experimental results that we compare against include reconstruction of the density field to better recover the linear power spectrum (dating back to Peebles \cite{peebles_tracing_1989} and Eisenstein et al. \cite{eisenstein_improving_2007}), throughout this paper we need to include it in our Fisher-based analyses as well. Therefore, as an estimation of the reconstruction, we decrease $\Sigma_{\parallel}$ and $\Sigma_{\perp}$ by 50\% following ref.\ \cite{seo_improved_2007} and \cite{eisenstein_improving_2007}.

\begin{center}

\begin{table}[t]
\caption{\label{tab:summary} A list of fiducial cosmologies used for each survey in the Fisher analysis.\\}
\centering
 \begin{tabular}{||c | c c c c c c ||} 
 \hline
 Survey & $\Omega_m$	& $\Omega_b$ & $h$ & $\sigma_8$ & $n_s$ & $\Omega_{\nu}$  \\ [0.5ex] 
 \hline\hline

 6dFGS &0.3 &0.0478 &0.70 &0.82 &0.96 &  0 \\
 \hline
 MGS	&0.31	&0.048	&0.67	&0.83	&0.96	&0	 	\\
 \hline
 BOSS (DR12)	&0.31	&0.04814	&0.676	&0.8	&0.97	&0	 \\
 \hline
 BOSS (DR9-11)	&0.274	&0.0457	&0.70	&0.8	&0.95	&0	 \\
 \hline
 eBOSS	&0.31	&0.04814	&0.676	&0.8	&0.97	&0.0014	 \\
 \hline
 WiggleZ &0.27 &0.04483 &0.71 &0.8 &0.963 &0  \\ 
 \hline
\end{tabular}
\end{table}
\end{center}

\subsection{The Fisher matrix for the RSD measurements}\label{subsec:Fisher_RSD}

In redshift space, the clustering of galaxies is distorted along the line of sight due to peculiar velocities. Measuring these redshift-space distortions (RSD) can provide a estimate of the growth rate of structure. In this Section, we describe how we predict such constraints using the Fisher formalism, following the method described in White et al. \cite{white_forecasting_2009}. To start with, we consider $\{p_1, p_2\} = \{\ln{b \sigma_8}, \ln{f\sigma_8}\}$ as the set of our free parameters. The parameter of interest constraining the structure growth is $f\sigma_8$, where $f$ is the logarithmic growth rate, and $\sigma_{8}$ is the amplitude of fluctuations in an $8 \mpcoh$ radius. For the purpose of this paper the galaxy bias, $b$, is a nuisance parameter over which we marginalize.

We can rewrite Eq.~\ref{eq:powerspectrum} as
\begin{equation}
    P = \big(b\sigma_8(z)+f\sigma_8(z)\mu^2\big)^2 \frac{P_m(k, z)}{\sigma_8(z)^2}\,.
    \label{eq:powerspectrum2}
\end{equation}
Then after taking the partial derivatives we obtain
\begin{equation}\label{eq:partial_RSD}
    \begin{aligned}
        \frac{\partial \ln{P}}{\partial \ln{p_1}} = \frac{2b\sigma_8(z)}{b\sigma_8(z) + f\sigma_8(z) \mu^2}\,, \\
        \frac{\partial \ln{P}}{\partial \ln{p_2}} = \frac{2\mu^2 f \sigma_8(z)}{b\sigma_8(z) + f\sigma_8(z) \mu^2} \,.
    \end{aligned}
\end{equation}
By inserting Eq.~\ref{eq:partial_RSD} into Eq.~\ref{eq:Fisher3}, we can derive the constraints on $f\sigma_8$. In order to provide consistent predictions for all surveys for our baseline RSD-based Fisher predictions we assume that the dilation parameters $\alpha_{\parallel}$ and $\alpha_{\perp}$ are held fixed, rather than marginalizing over them. This limits the dependence on the BAO detection, which in turn controls how well the dilation parameters are constrained. Therefore, our results should not be directly compared to those from measurements where they marginalize over these parameters after performing a joint fit to data. 

For many of the surveys, the authors provide errors for both fixed BAO dilation parameters (``$\alpha$s'') and results after marginalizing over them, which we refer to as $f\sigma_8^{\rm fx. \, \alpha s}$, and $f\sigma_8^{\rm mg. \, \alpha s}$, respectively. Other analyses provided the covariance matrices for the four parameters, from which we can calculate both.
When marginalizing over dilation parameters, we simply take the square root of the $f \sigma_8$ diagonal element in the covariance matrix, $f\sigma_8^{\rm mg. \, \alpha s} = \sqrt{C_{f\sigma_8, f \sigma_8}}$. To find $f\sigma_8^{\rm fx. \, \alpha s}$, we calculate the $f\sigma_8$ error from the $f \sigma_8$ diagonal element of the inverse covariance matrix (the survey's Fisher matrix), $f\sigma_8^{\rm fx. \, \alpha s} = \left[C^{-1}\right]_{f\sigma_8, f \sigma_8}^{-1/2}$. In Section~\ref{sec:Data}, we briefly describe which method is used to set the correct values in Table~\ref{tab:results} and Table~\ref{tab:results2}.

\subsection{Integration Limits ($k_{\rm min}$ and $k_{\rm max}$)}\label{subsec:kmax}

To forecast the Fisher-based analysis parameters, we need to make assumptions about the upper and lower limits of the integral in Eq.~\ref{eq:Fisher2}. Each survey is able to extract information up to scales comparable to its size. Formally, the integral constraint affects the power spectrum such that a copy of the window function centred at $k=0$ is subtracted from the convolved power, leaving zero power at $k=0$ (e.g. \cite{beutler_clustering_2017}). Thus, information on scales the size of the survey window is not present. To match this behaviour, we choose the lower limit of our integral over scales to be $k_{\rm min} = 2\pi V_{\rm sur}^{-1/3}$ (\cite{tegmark_measuring_1997, zhao_extended_2016}) mimicking the effect of the window with a sharp cut in scales included. The choice of $k_{\rm max}$, on the other hand, depends on the scales to which we can extract linear information. The non-linear evolution primarily affects the BAO through well controlled damping terms $\Sigma_{\perp}$ and $\Sigma_{\parallel}$ , and thus Seo \& Eisenstein \cite{seo_improved_2007} suggested that when extracting the BAO parameters, we can set $k_{\rm max}=0.5 \hompc$. Indeed, we find that $\sigma_{\ln{D_V}}$ is not sensitive to the choice of $k_{\rm max}$ for $k\gtrsim0.3 \hompc$; the BAO provide a large-scale signal localized in configuration space, such that in $k$-space the signal rapidly diminishes to small scales (See the left panel of Fig.~\ref{fig:Errors_vs_kmax}). This finding agrees very well with results from N-body simulations \cite{seo_improved_2007}, which also found that the error is stable for $k_{\rm max} = 0.3, 0.4$, and $0.5 \hompc$.

For the RSD measurements, we found that $\sigma_{\ln{f \sigma_8}}$ is highly dependent on the choice of $k_{\rm max}$. On small scales, the density field becomes highly non-Gaussian and hence the inverse of the linear Fisher matrix gives a more optimistic estimation of the error bars than the measurements. The reduction in linear information available is gradual: in models, this results in an increased dependence on non-linear parameters, often allowed to be free given unknown non-linear effects including beyond-linear galaxy bias. The exact scale at which we stop being able to recover linear information is expected to depend on the details of the galaxy population, and on the accuracy and number of free parameters included in the model used. This problem leads us to test the Fisher error bars' sensitivity to the change of $k_{\rm max}$. Our default prediction is to calculate the expected error on $f \sigma_8$ up to a fiducial $k_{\rm max} = 0.1 / D(z = z_{i}) \hompc$ at each redshift slice. This is based on arguments made by Okumura et al. \cite{okumura_distribution_2012}, where they showed that after this scale the power spectrum turns strongly non-linear to the extent that a Taylor series cannot adequately describe the redshift-space density field anymore \cite{font-ribera_desi_2014}. 

We also consider inverting the problem and using the data measurements to determine what $k_{\rm max}$ we should use. To do so, we vary $k_{\rm max}$ from $0.01/D(z_i)$ up to $0.5/D(z_i)$ and we plot $f \sigma_8$ error against $k_{\rm max}(z_{\rm eff})$ in Figure~\ref{fig:Errors_vs_kmax}. This allows us to translate the constraining power of RSD measurements to an effective $k_{\rm max}$ at which an equivalent amount of information can be extracted from the linear power spectrum.

\section{Data and modelling} \label{sec:Data}

In this section, we introduce the surveys on which we perform our Fisher analysis: SDSS-I\&II Main Galaxy Sample (MGS), SDSS-III Baryon Oscillation Spectroscopic Survey (BOSS), SDSS-IV extended Baryon Oscillation Spectroscopic Survey (eBOSS), and WiggleZ Dark Energy Survey (WiggleZ). To test our code, we compare to the predictions in Zhao et al. \cite{zhao_extended_2016} for eBOSS, and find that both codes match in giving very similar results for the same input parameters. 
In our paper, we are only interested in understanding and matching to the statistical errors and so, where appropriate, we have removed the quoted systematic errors from any combined constraints by subtracting them in quadrature. 

Uncertainty in redshift estimation can increase uncertainty in the BAO and RSD features by damping the radial component of the power spectrum (as explained in Section~\ref{sec:method}). The observations require that $\sigma_v \equiv c \sigma_z / (1+z) \leq 10^{-3} c$ for galaxies at all redshifts. However, the quasar clustering measurements suggest that while this requirement holds true for low redshifts, higher redshifts are prone to higher velocity errors. Therefore, when performing our Fisher analysis on the surveys described in Section~\ref{sec:Data}, we assume that the velocity error for quasars has the following form (Zarrouk et al. \cite{zarrouk_clustering_2018}): 
\begin{equation}\label{eq:velocity_error}
    \frac{\sigma_{v}}{c}= \left\{\begin{array}{lr}
        10^{-3}, & \text{if } z\leq 1.5\\
        \frac{4}{3} \times 10^{-3}(z-1.5)+10^{-3}, & \text{if } z > 1.5
        \end{array}\right\}\,.
\end{equation}
We use $\sigma_v = 10^{-3}c$ for all other tracers,
and we include the redshift error in both the RSD
and BAO Fisher predictions.

Many of the papers introduced in the following sections use mock catalogues for the following reasons: to estimate systematic errors, to test the model of power spectrum or the correlation function, and to better estimate the covariance matrices from the measurements. Noise in the data may lead to error bars that are larger or smaller than the true constraining power of the survey just by chance. To examine this possibility, we also compare the  $f \sigma_8$ constraints on mocks to our Fisher-based errors, where available. In general, the results from the fits to mocks provide error bars similar to those obtained from the data (results are shown in left panel of Figure~\ref{fig:kmax_vs_zeff}).

\subsection{Reconstruction Technique}

As already described, the BAO feature can be estimated either by measuring the peak in the correlation function or by the harmonic sequence of oscillations in the power spectrum. As gravitational forces make structure grow through time, this signature blurs, meaning that the precision at which we can measure the BAO signal decreases. In order to sharpen the broadened BAO signal, various reconstruction methods have been proposed based on the idea of rewinding the motion of galaxies to move them into their original positions. The strong impact of using even simple methods to do this on BAO measurements was first described by Eisenstein et al.\ (\cite{eisenstein_improving_2007,eisenstein_robustness_2007}),  who proposed a method to shift the galaxies' position by the linear-theory estimated Lagrangian displacement field and showed that this process can increase the precision of the BAO signal. In this section, we discuss some reconstruction methods that have been applied to the data papers that we consider throughout this paper.

A significant difficulty in implementing reconstruction arises because of the RSD, and particularly that the RSD direction changes across a survey. This problem has been solved in recent analyses using two different methods. Padmanabhan et al.\ \cite{padmanabhan_reconstructing_2009} implemented a finite-difference routine based solving for the potential on a grid covering the survey's volume. The direction of the RSD signal is allowed to vary for different grid points. Burden et al.\ \cite{burden_efficient_2014,burden_reconstruction_2015} instead showed how the simple linear theory based reconstruction method can be undertaken in the Fourier space. To allow for RSD without forcing a global plane-parallel approximation, the code is iterative, removing the estimated RSD signal based on the potential field found in a previous step where no RSD was assumed to be present. The authors also tested their method on CMASS DR11 mocks and found that it converges rapidly, requiring only two iterations.

The way in which reconstruction works is described in more detail in Padmanabhan et al.\ \cite{padmanabhan_2_2012} and one important aspect is that the field be smoothed, with a smoothing scale of between 10 and 15 $\mpcoh$ \cite{burden_reconstruction_2015}. Note that while many techniques also leave a field in which the RSD have approximately been removed, this is not a necessary part of the code and the BAO positions and signal strength do not depend on this. The power spectrum (or correlation function) after reconstruction has a complicated form: ref.~\cite{padmanabhan_2_2012} suggested that the power spectrum be modeled using three nonlinear damping terms and over three wavelength ranges. Thus, while modelling the full post-reconstruction clustering signal is difficult and may change the shape of the clustering in a hard-to-model way, what is clear is that these reconstruction techniques can decrease $\Sigma_{\parallel}$ and $\Sigma_{\perp}$ by 50\% for the BAO. Thus if one is only concerned with the BAO signal, allowing for smooth changes in the shape of the power to isolate only this signal removes the pernicious effects of reconstruction. Applying this method to the SDSS DR7 sample showed that it reduces the BAO distance error by a factor of 1.8 \cite{padmanabhan_2_2012}. The improvement is not universal, with samples reacting differently to reconstruction \cite{anderson_clustering_2012}, but it clearly works to improve the average recovered signal for a set of volumes of the Universe.

The methods described above are relatively simple, relying only on linear physics. More sophisticated techniques offer the promise of increased improvements in the future (e.g.\ \cite{birkin_reconstructing_2019, sarpa_bao_2019, modi_cosmological_2018, nikakhtar_laguerre_2021}). All of the analyses we consider in this paper, except for quasars, used simple methods for reconstruction, and we assume a 50\% improvement on $\Sigma_{\parallel}$ and $\Sigma_{\perp}$ from this.

\subsection{RSD modelling}

The amount of information that RSD surveys can extract is limited by  modelling the power spectrum or correlation function: the data, at least in 2-point form, does not itself provide information about which scales are linear. Even if the data match a linear model, this does not mean that all of the linear information is present, as cancellation of multiple effects is possible (e.g.\ Fingers-of-God and non-linear growth in the monopole). Therefore, the fidelity of the RSD modelling will limit the amount of information extracted. In this section we briefly describe the models used in the data papers that we consider.

\emph{Fourier space:} 
The simplest RSD model for the power spectrum was first introduced by Kaiser in 1987 \cite{kaiser_clustering_1987}. As it does not include nonlinearities in the halo power spectrum, it is therefore only applicable on large scales. Scoccimarro later constructed a fitting model for RSD in 2004 (Sc.; \cite{scoccimarro_redshift-space_2004}), as a nonlinear extension to the linear Kaiser model (1987) \cite{kaiser_clustering_1987} with two free parameters. Scoccimarro's extension can include Gaussian and non-Gaussian contributions to the velocity dispersion of large-scale flows. Later in 2010, the matter power spectrum in redshift space was developed by Taruya, Nishimichi, and Saito (TNS; \cite{taruya_baryon_2010}). They added various coefficients to the Scoccimarro's model to account for nonlinearities between the density field and the velocity field and presented a new power spectrum in redshift space for modeling BAO, including nonlinear gravitational clustering and RSD. The TNS model is amongst the most popular RSD models in Fourier space, of which many of the most recent surveys described in this section make use. Since all of these fitting formulae break down on small scales, they limit their analyses to a maximum $k$, which is reported as $k_{\rm max}(\rm{O.})$ in Table~\ref{tab:results} and~\ref{tab:results2}.

\emph{Configuration space:} 
One model that includes nonlinearities in the  correlation function in redshift space at quasi-linear scales is the streaming model by Reid \& White \cite{reid_towards_2011}. They modelled the nonlinear mapping between the real and redshift space with the Gaussian streaming model, in which they included the dependence of halo pairwise velocities on their separation and angle with respect to the line-of-sight. We refer to this RSD model as R+11.

A more realistic way to model RSD in configuration space is to model the intrinsic galaxy clustering and the velocity field with Convolved Lagrangian Perturbation Theory (CLPT; \cite{wang_analytic_2014}) and then model the convolution of the velocity field along the line-of-sight with the Gaussian Streaming model (GS; \cite{reid_towards_2011}). Throughout, we refer to this method as CLPT+GS RSD model.
On the other hand, Jennings et al. (J+11; \cite{jennings_modelling_2010}) proposed a cosmology-independent relationship between the velocity field and the density field, from which they found an RSD fitting model based on the nonlinear velocity divergence matter power spectrum.
In addition, Sanchez et al. \cite{sanchez_clustering_2013} described a simple recipe for modelling the full shape of the clustering wedges that we refer to as S+13. 
The nonlinear power spectrum in this model is motivated by RPT \cite{crocce_renormalized_2006}.
A variation of this method, gRPT \cite{sanchez_clustering_2017,Pezzotta21} has also been used
by ref.~\cite{grieb_clustering_2017}.

The maximum wavenumber used in Fourier analyses, as stated before, is given by the so-called $k_{\rm max}$. The analogous scale in real space is the smallest scale in the correlation function from which authors can extract information, and is usually referred to as $ s_{\rm min}$. In our paper, we are interested in comparing the Fisher-based $k_{\rm max}$ with that of the measurements. For the purposes of Table~\ref{tab:results} and Table~\ref{tab:results2}, whenever the analysis is done in configuration space, we approximate $k_{\rm max}$ with $ \sim 1.15\pi/s_{\rm min}$, based on arguments made by \cite{reid_towards_2011}. This allows us to compare the smallest scales used in both real and Fourier space RSD analyses with the effective $k_{\rm max}$ corresponding to the total amount of linear information available.

\subsection{6-degree Field Galaxy Survey}

The 6dFGS survey was undertaken from 2001 to 2006 using the Six-Degree Field multi-fibre instrument of the UK Schmidt Telescope (UKST) \cite{jones_6df_2004}. This survey covered more than $125,000$ galaxies over $\sim 17,000$ deg$^2$ of the southern sky in the redshift range $z < 0.3$ with a median redshift of $z_{\rm med} =  0.053$.

For the BAO analysis, we use the same catalogue as Carter et al. \cite{carter_low_2018}, which contains $75,117$ galaxies after applying cuts to the magnitude and the completeness. This catalogue has an effective redshift of $z_{\rm eff} = 0.097$ and an effective bias of $b_{\rm eff} = 1.65$. The early analysis of Beutler et al. \cite{beutler_6df_2011} was recently supplanted by an analysis that used more modern techniques, including reconstruction from Burden et al. (2014, 2015) with a smoothing scale of $15\mpcoh$ and a covariance matrix based on more sophisticated simulations (i.e. \cite{carter_low_2018}). They found that the 6dFGS likelihood is bimodal, with a 4.6\% error on $D_V$ for the best fit model. The authors combined the post-reconstruction 6dFGS with the SDSS MGS sample and reported the lowest fractional error to date on $D_V$, 3.2\%, at low redshift. This favoured the second most likely peak seen when fitting the post-reconstruction BAO signal in the 6dFGS. The 6dFGS sample only adds enough information to provide an improvement of $\sim16$\% on the MGS BAO measurements at low redshift. For our analysis we include results from the 6dFGS separately from the MGS as we are interested in the surveys independently.

Beutler et al. in 2012 \cite{beutler_6df_2012}, used a slightly different catalogue containing $81,971$ galaxies to make RSD-based measurements. They made use of two RSD models for the 2D correlation function in configuration space, namely, the Simple Streaming model down to $r = 10 \mpcoh$ and the Scoccimarro model down to  $r = 16 \mpcoh$ or $k_{\rm max} \sim 0.23 \hompc$ (Sc. in Table~\ref{tab:results}). We only report the results with the Scoccimarro model as it only fits for the two parameters of interest, $f \sigma_8$ and $b \sigma_8$, and it gives a fractional error on their measurement of $f\sigma_8^{\rm fx. \, \alpha s}$ of 13.0\%. 
Since they do not fit the dilation parameters in their model, we consider their result as if dilation parameters were fixed, i.e, the error on $f\sigma_8^{\rm fx. \, \alpha s}$ is 13.0\%, and in Table~\ref{tab:results} we do not report the $f\sigma_8^{\rm mg. \, \alpha s}$ for this survey. They also varied their fiducial cosmology without their results changing. This is because the degeneracy between RSD parameters and dilation parameters is very small at low redshift.

\subsection{Main Galaxy Sample SDSS-I\&II}
The Main Galaxy Sample (MGS) is a part of the seventh data release (DR7; \cite{abazajian_seventh_2009}) of SDSS I\&II (\cite{york_sloan_2000}), using observations from the 2.5-meter Sloan telescope located at Apache Point Observatory (APO; Gunn et al. \cite{gunn_25_2006}). In this paper, we consider the subsample of $63,163$ galaxies covering $6,318\ \rm deg^2$ of the sky, with a redshift range of $0.07 < z < 0.2$ and an effective bias of 1.5 created by Ross et al.\ \cite{ross_clustering_2015}. The MGS sample contains significantly more galaxies than this but, because it was volume limited, a high bias subsample was selected for analysis in order to facilitate the creation of mock catalogues from simulations (which then only required halos to a higher halo mass limit). Ross et al. followed the standard linear reconstruction prescription using the Fourier based method, and found that reconstruction improved the BAO signal by a factor of 2. The post-reconstruction $D_V$ was measured to an accuracy of $3.8\%$, and there was no evidence of systematic errors. In addition, based on the post-reconstruction BAO measurements and using the CLPT RSD model in range $25 < s < 160 \mpcoh$ ($k_{\rm max} = 0.14 \hompc$), Howlett et al. \cite{howlett_clustering_2015} measured $f\sigma_8^{\rm fx. \, \alpha s}$ and $f\sigma_8^{\rm mg. \, \alpha s}$ to an accuracy of 31.8\% and 40.5\% respectively (Table~2, eighth and second case).\footnote{Since the error bars in this analysis are asymmetric, we average the upper and lower error bars.}
Moreover, in order to estimate the covariance matrix precisely, 1000 mock catalogues from PICOLA code have been analysed for the MGS sample by Howlett et al. \cite{howlett_clustering_2015}. They found that the best fit value from the average of these mocks gives $f\sigma_8^{\rm fx. \, \alpha s} = 0.5^{+0.13}_{-0.12}$. (Table~1, 8th case).

\subsection{Baryon Oscillation Spectroscopic Survey SDSS-III}
The Baryon Oscillation Spectroscopic Survey (BOSS) \cite{dawson_baryon_2013} is a part of the SDSS-III \cite{eisenstein_sdss-iii_2011} that was undertaken by the 2.5-meter Sloan Telescope from 2008 to 2014. Covering an area of 10,000 deg$^2$, BOSS contains more than 1.5 million galaxies with redshifts up to $z=0.7$. Two target selection algorithms were used to create the BOSS galaxy sample: LOWZ (at lower redshift) is a selection of luminous red galaxies to $z \approx 0.4$, and CMASS (for constant stellar Mass) covers LRGs up to a higher redshift range $z \lesssim 0.7$ \cite{reid_sdss-iii_2015}. We review the catalogue selection and the BAO and RSD measurements in three early data releases of BOSS, DR9, DR10, and DR11 in Section~\ref{subsec:DR9-11}, and the final data release DR12, in Section~\ref{subsec:DR12}.

\subsubsection{Intermediate Data Releases DR9-11}\label{subsec:DR9-11}

The BOSS DR9 CMASS sample contains $\rm 264,283$ galaxies in a region of $3,275 \rm deg^2$ of the sky. Galaxies used in this catalogue cover a redshift range of $0.43<z<0.7$ with an effective redshift of $z_{\rm eff} = 0.57$. BAO measurements of were presented by Anderson et al.\ \cite{anderson_clustering_2012} and showed that $\sigma \ln{D_V} = 1.6\%$. Anderson et al.\ used the finite difference reconstruction method with a smoothing scale of $15 \mpcoh$, and found that applying reconstruction to this particular data does not improve the precision of the BAO feature. They suggested that this is because the pre-reconstruction errors of this sample were already at the lower end of the expected range (from mocks)---hence there was little for reconstruction to improve. RSD measurements for this survey were presented in Reid et al. \cite{reid_clustering_2012} providing fractional error bars on the growth rate $f\sigma_8^{\rm mg.\,\alpha s} = 14.6 \%$ and $f\sigma_8^{\rm fx.\,\alpha s} = 8.1 \%$. For both of these analyses, the systematic errors were negligible compared to the statistical errors. Their findings show that $b\sigma_8(z_{\rm eff}) = 1.2$, where $\sigma_8(z_{\rm eff}) = 0.61$, which gives an effective bias of $2.0$. They fit the monopole and quadrupole moments of the correlation function down to scales of $s_{\rm min}=25\mpcoh$ (or $k_{\rm max} \sim 0.14\hompc$) with Reid \& White's RSD model \cite{reid_towards_2011}.

BOSS Data Release 10 (DR10) contains $218,905$ galaxies in LOWZ ($0.15<z<0.43$) and $501,844$ galaxies in CMASS ($0.43<z<0.7$). Using these catalogues, $D_V$ was constrained to 2.8\% and 1.4\% for LOWZ and CMASS, respectively, by Anderson et al.\ \cite{anderson_clustering_2014}. They also measured the anisotropy distances for CMASS in this catalogue and found a consensus ($P(k)+ \xi(s)$) error of 1.9\% and 5.0\% on $D_M$ and $D_H$, respectively. Anderson et al. \cite{anderson_clustering_2014} also analyzed the 11th data release of BOSS, which consists of  $313,780$ galaxies in LOWZ and $690,826$  galaxies in CMASS. This gave a statistical consensus ($P(k)+ \xi(s)$) error of of 2.0\% and 0.9\% on $D_V$ for LOWZ and CMASS, respectively (after subtracting 0.3\% systematic error in quadrature). They also found errors of 1.4\% and 3.5\% for the anisotropic distances, $D_M$ and $D_H$, respectively. For both DR10 and DR11 data, the reconstruction method by Padmanabhan et al. was applied to NGC and SGC separately.

Sanchez et al.\ \cite{sanchez_clustering_2014} constrained the logarithmic growth of structure using the 10th and 11th BOSS data releases, using the S+13 model described earlier, within the range of $40\mpcoh<s<160\mpcoh$ or a $k_{\rm max}$ of $0.09\hompc$. They found fractional errors of 23.3\%, 12.8\%, 20.8\%, and 10.8\% on $f\sigma_8^{\rm mg. \, \alpha s}$, for LOWZ DR10, CMASS DR10, LOWZ DR11, and CMASS DR11, respectively. 

Since Sanchez et al.\ did not publish either the $f\sigma_8^{\rm fx. \, \alpha s}$, nor the covariance matrix including growth rate and dilation constraints for BOSS DR10 and DR11 samples, we are limited in how well we can replicate these results. Therefore, we use the CMASS DR11 RSD measurements from Samushia et al. \cite{samushia_clustering_2014} instead. These used the RSD streaming model described in Reid \& White (R+11, \cite{reid_towards_2011}) with $s_{\rm min} = 25 \mpcoh$ or $k_{\rm max} \sim 0.14 \hompc$, finding errors of 9.9\% and 6.0\% on $f\sigma_8^{\rm mg. \, \alpha s}$ and $f\sigma_8^{\rm fx. \, \alpha s}$ respectively, which we include in Table~\ref{tab:results2}. Systematic errors have been ignored in this analysis since they have been checked with mocks and they had less than a 1\% effect. When fitted to growth of structure separate from the dilation parameters, they found $b \sigma_8 = 1.26$, which gives an effective bias of $b(z_{\rm eff}) = 2.05$ given that $\sigma_8(z = 0.57) = 0.615$. 
 

\subsubsection{Final Data Release DR12}\label{subsec:DR12}
To determine the Fisher-based predictions for the final data release of BOSS, DR12, we follow the data selection method of two studies. First, Alam et al. \cite{alam_clustering_2017} who split this sample by redshift range (Near and Mid in Table~\ref{tab:results}) and second, Gil-Marín et al. \cite{gil-marin_clustering_2016}, who studied the LOWZ and CMASS catalogues separately (LZ and CM in Table~\ref{tab:results2}). In the following, we discuss the recovered errors for BAO and RSD measurements for both selection methods.

Alam et al.\ combined LOWZ with CMASS, and after applying redshift cuts to the combined sample, they created three partially overlapping redshift samples that cover 9329 deg$^2$ area of the sky.
In this paper, we refer to the first redshift bin, at $0.2 < z < 0.5$, which mainly consists of the LOWZ galaxy sample, as the near redshift bin, and the second redshift bin at $0.4 < z < 0.6$, mainly consisting of the CMASS sample, as the mid redshift bin, with effective redshifts $z_{\rm eff,1} = 0.38$ and $z_{\rm eff,2} = 0.51$ respectively. In our work, we do not consider the higher redshift bin at $0.5<z<0.75$ of the BOSS survey, as it is combined with the eBOSS LRG samples (refer to Section~\ref{subsec:LRG}). 
According to Table~3 in Beutler et al., BOSS has an effective bias of 2.03, and 2.13 in the LOWZ and CMASS samples. Note that the Luminous Red Galaxy (LRG) samples in this paper are assumed to have a galaxy bias of $b_{\rm LRG}(z) = 1.7 / D(z)$, where D(z) is the linear growth factor. This assumption is consistent with the fiducial bias assumed in Zhao et al. \cite{zhao_completed_2020} and Prakash et al. \cite{prakash_sdss-iv_2016}, as well as the effective biases measured for the LOWZ and CMASS samples. 

The post-reconstruction BAO-only analysis by Alam et al.\ yields a 1.5\%, 2.7\%, and 1.0\% statistical uncertainty in $D_M$, $D_H$, $D_V$ for the BOSS Near sample. For the BOSS Mid sample, these uncertainties are lower: 1.4\%, 2.3\% and 0.9\% for $D_M$, $D_H$, and $D_V$ respectively. 
They used the reconstruction method described in Padmanabhan et al.
For the constraint on the growth rate, Alam et al. incorporated results from 4 different papers, using different methods: real-space multipoles (Satpathy et al. \cite{satpathy_clustering_2017}; CLPT+GS, $s_{\rm min} = 25 \mpcoh$),
real-space wedges 
(Sanchez et al. \cite{sanchez_clustering_2017}; similar to TNS, $s_{\rm min} = 20 \mpcoh$),
Fourier-space multpoles (Beutler et al. \cite{beutler_clustering_2017}; TNS, $k_{\rm max} = 0.15 \hompc$),
and Fourier-space wedges (Grieb et al. \cite{grieb_clustering_2017}; gRPT+RSD, $k_{\rm max} = 0.2 \hompc$). The BAO+FS consensus measurements from all of these works yield a 7.8\% statistical error on $f\sigma_8^{\rm mg. \, \alpha s}$ for the near redshift bin, and 7.6\% for the mid redshift bin. We utilized the covariance matrix to obtain the constraint on $f\sigma_8^{\rm fx. \, \alpha s}$, which is 7.0\% for the near redshift slice and 6.4\% for the mid redshift slice.

In addition, BOSS DR12 measured BAO (\cite{gil-marin_clustering_2016-2}) and RSD (\cite{gil-marin_clustering_2016}) using LOWZ and CMASS samples, consisting of $361,762$, and $777,202$ galaxies within the redshift ranges $0.15 < z < 0.43$, and $0.43 < z < 0.70$, and with effective redshifts of 0.32, and 0.57. They found that $b \sigma_8$ for these two samples are 1.29 and 1.24, resulting in an effective bias of 1.9 and 2.1, respectively. Throughout their study, they have made use of the finite-difference reconstruction method by Padmanabhan et al.

The BAO-only consensus analysis in real space and Fourier space have shown errors of 2.2\%, 5.9\%, and 1.7\% on $D_M$, $D_H$, and $D_V$ in the LOWZ sample, and 1.3\%, 2.9\%, and 0.9\% on $D_M$, $D_H$, and $D_V$ in the CMASS sample (Table~4 in \cite{gil-marin_clustering_2016-2}). Moreover, after modelling the RSD with the TNS model with a $k_{\rm max}$ of $0.24 \hompc$, Gil-Marín et al. \cite{gil-marin_clustering_2016} measured $f\sigma_8^{\rm mg. \, \alpha s}$ with an error of 15.7\% and 8.6\%, and when assuming no AP effect, they measured $f\sigma_8^{\rm fx. \, \alpha s}$ with an error of 9.1\% and 5.0\% in LOWZ and CMASS samples respectively (Table~3 in \cite{gil-marin_clustering_2016}).

\subsection{Extended Baryon Oscillation Spectroscopic Survey SDSS-IV}

The extended Baryon Oscillation Spectroscopic Survey (eBOSS; Dawson et al.) \cite{dawson_sdss-iv_2016} is a part of SDSS-IV (Blanton et al. \cite{blanton_sloan_2017}), and used the Sloan Telescope at Apache Point Observatory (APO; Gunn et al. \cite{gunn_25_2006}) to conduct a redshift survey at higher redshifts than BOSS from 2014 to 2019. From this survey's 16th data release, we use the DR16 Luminous Red Galaxy (LRG \cite{ross_completed_2020}; Section~\ref{subsec:LRG}), Emission Line Galaxy (ELG \cite{raichoor_completed_2020}; Section~\ref{subsec:ELG}), and Quasar samples (\cite{lyke_sloan_2020}; Section~\ref{subsec:QSO}), which are reported in Table~\ref{tab:results}. We have also used the LRG and Quasar samples from an earlier data release, DR14 (Pâris et al. \cite{paris_sloan_2018}), which are reported in Table~\ref{tab:results2}. Since for eBOSS the systematic errors are estimated using mock catalogues, in this section we only quote the statistical error bars.

\subsubsection{LRGs}\label{subsec:LRG}
There are $174,816$ LRGs in eBOSS DR16 sample that cover $9,493 \rm \,  deg^2$ of the night sky, with an effective redshift of $z_{\rm eff}  =0.698$. For LRGs in the redshift range $0.6<z<1.0$, we combine the eBOSS LRG sample with $202,642$ CMASS BOSS DR12 galaxies, to be consistent with the Bautista et al. \cite{bautista_completed_2020} contraints. 
We want the Fisher forecast to have the same effective bias as the full MCMC fit to the TNS model, which is $b_1 = 2.2$. Therefore, we assume that the linear bias has the form of $1.5/D(z)$.

The estimated covariance matrix reported by Bautista et al. \cite{bautista_completed_2020}, for the BAO-only analysis, after applying the reconstruction technique of Burden et al. \cite{burden_efficient_2014,burden_reconstruction_2015}, showed that the fractional statistical errors on $D_M$, $D_H$, and $D_V$, are $1.6\%$, $2.5\%$, and $1.6\%$ respectively. 

For obtaining the growth rate, they used CLPT+GS RSD to model correlation function in real space and combined their results with the Fourier-space analysis of Gil-Marín et al. \cite{gil-marin_completed_2020}, which used the TNS model. In this paper we use their consensus results. They tested for different fitting ranges of scales, and found that the optimal minimum scales that should be covered in the CLPT+GS and TNS model are $ 25 \mpcoh$ ($k_{\rm max} \sim 0.14 \hompc$) and $ 20 \mpcoh$ ($k_{\rm max} \sim 0.18 \hompc$), respectively. 

The consensus BAO + Full Shape RSD fit indicated that the error on $f\sigma_8^{\rm mg. \, \alpha s}$ and $f\sigma_8^{\rm fx. \, \alpha s}$ are 9.4\% and 9.1\% (from eq.~56 in ref.~\cite{bautista_completed_2020}). In Table~10 of Bautista et al. they evaluated the systematic errors using mocks, and showed that the systematic error of $f\sigma_8$ for CLPT and TNS is 0.024 and 0.023 respectively. If we subtract $\rm \sim 0.0235$ from the total error obtained from Eq.~56, in quadrature, then we obtain an error of 7.9\% and 7.6\% on $f\sigma_8^{\rm mg. \, \alpha s}$ and $f\sigma_8^{\rm fx. \, \alpha s}$. Furthermore, for this catalogue, they constructed 1000 EZmock realisations for the LRG eBOSS+CMASS survey geometry and obtained 9.9\% error on $f\sigma_8^{\rm fx. \, \alpha s}$ in the mocks, which is shown in Figure~\ref{fig:kmax_vs_zeff}.


An earlier data release of the eBOSS LRG sample, DR14, when combined with CMASS, is comprised of $126,557$ galaxies in redshift range $0.6 < z < 1.0$, with an effective redshift $z_{\rm eff} = 0.720$. Bautista et al. \cite{bautista_sdss-iv_2018} followed the reconstruction technique presented by Burden et al. \cite{burden_efficient_2014,burden_reconstruction_2015}, and the post-reconstruction BAO measurement analysed by them showed a 2.5\% fractional error on $D_V$. The anisotropic fits on this data shows very different upper and lower error bars: $D_M = 2689_{-79}^{+158}$, and $D_H = 2593_{-589}^{+177}$. Hence we do not report these errors in Table~\ref{tab:results2}. But the isotropic fit shows that $D_V = 2377_{-59}^{+61}$ which translates into 2.5\% fractional error.
The RSD measurement for this sample has been modeled with CLPT+GS model by Icaza-Lizaola et al. \cite{icaza-lizaola_clustering_2020}
on scales $28 < s < 124$ $\hompc$ ($k_{\rm max} = 0.13\hompc$) and they found a fractional error of 29.2\% on $f\sigma_8^{\rm mg. \, \alpha s}$. Using the covariance matrix of their analysis, we found a fractional error of 10.4\% for $f\sigma_8^{\rm fx. \, \alpha s}$. Their fit showed that $b \sigma_8 = 1.11$ (Table~B1 in ref.~\cite{icaza-lizaola_clustering_2020}), which means assuming $D(z_{\rm eff} = 0.72) = 0.691$, the effective bias would be $b_{\rm eff} = 2.0$. We therefore use the functional form of $1.4 / D(z)$ for bias evolution in our Fisher analysis.

\begin{figure}[t]
  \centering
  \includegraphics[width=1\textwidth,angle=0]{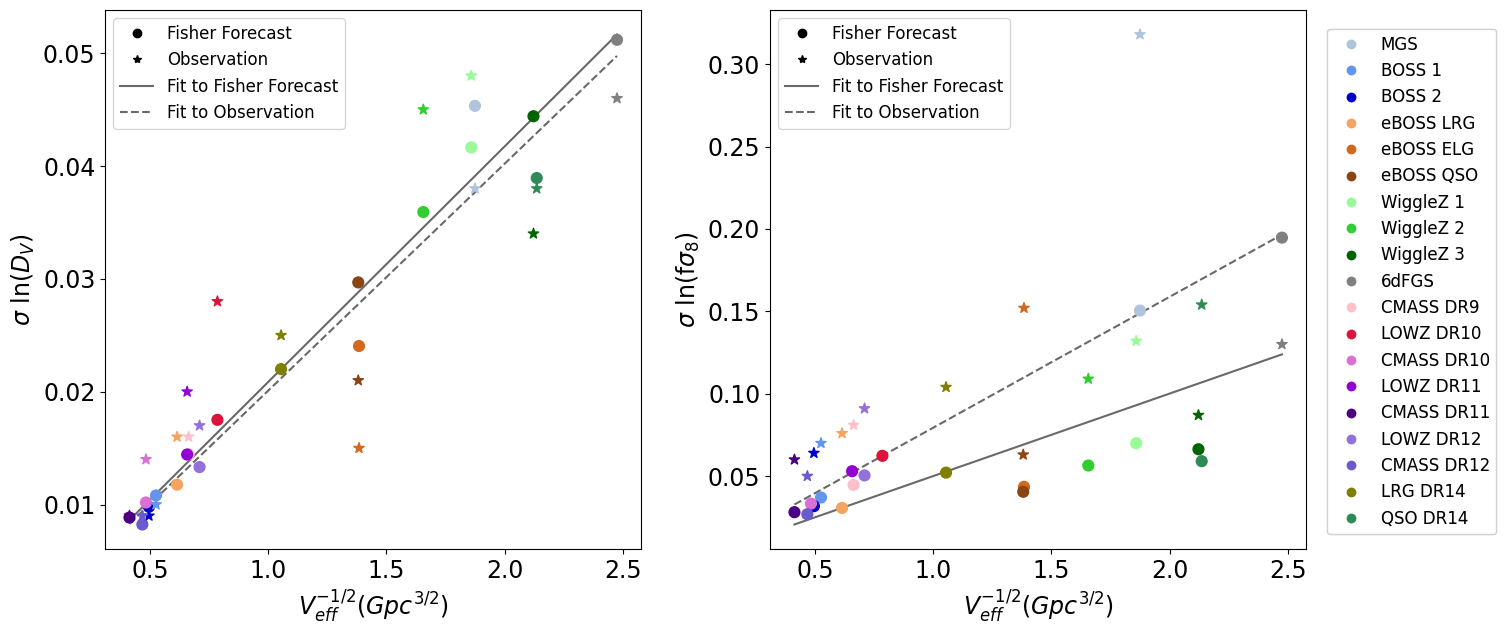}
  
  \caption{$D_V$ fractional error as measured from the BAO position plotted against the effective volume $V_{\rm eff}^{-1/2}$ for different surveys (left panel). A clear trend is seen, with the slopes of the linear fit to the Fisher and observational errors being 0.022 and 0.020, respectively. The measured RSD constraints show a weaker correlation with $V_{\rm eff}^{-1/2}$.
  }
  \label{fig:Errors_vs_Veff}
\end{figure}

\subsubsection{ELGs}\label{subsec:ELG}
The $16^{th}$ data release of SDSS-IV contains $173,736$ Emission Line Galaxies (ELGs) covering an effective area of 727 $\rm deg^2$ at effective redshift $z_{\rm eff}  = 0.845$ in the redshift range $0.6 < z < 1.1$ (de Mattia et al. \cite{de_mattia_completed_2020} \& Tamone et al. \cite{tamone_completed_2020}). Tamone et al. fitted for galaxy bias and found that $b = 1.52$. Therefore, for this sample, we assume a galaxy bias function of $b_{\rm ELG}(z) = 0.99/ D(z)$, as $D(z_{\rm eff} = 0.85) = 0.651$. 

De Mattia et al. applied the iterative FFT reconstruction method introduced by Burden et al. and assumed a Gaussian smoothing scale of 15 $\mpcoh$.
The isotropic BAO measurements in Fourier space for this sample gave statistical upper error of 2.5\% and lower error of 2.8\% for $D_V$, which is a factor of 1.2 smaller than the total errors. We averaged the upper and lower error bars for reporting in Table~\ref{tab:results} (refer to Table~9 of ref. \cite{de_mattia_completed_2020} for statistical-only errors). 

\begin{figure}[t]
    \centering 
    \includegraphics[width=1\textwidth,angle=0]{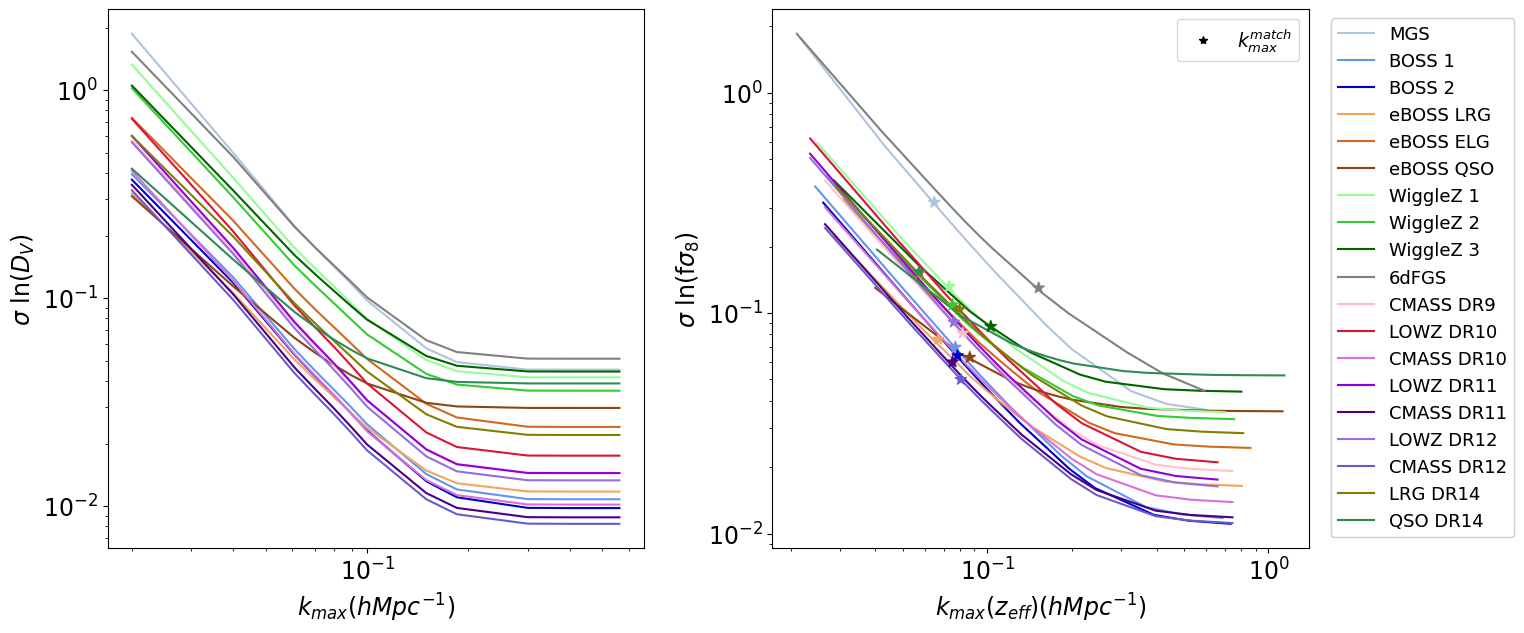}
    \caption{
    The Fisher-based predicted fractional error on $D_{V}$ (\emph{left}) and $f\sigma_8$ (\emph{right}) plotted against $k_{\rm max}(z = z_{\rm eff}$) for different surveys. 
    Note that in contrast to the $f \sigma_8$ error, the $D_{V}$ error converges to a constant at $k \gtrsim 0.3 \hompc$.
    The stars on the right panel indicate the $k_{\rm max}$ at which the Fisher prediction and the experiment are equal.
    }
    
    \label{fig:Errors_vs_kmax}
\end{figure}

They also performed the Fourier space BAO+RSD measurements using the TNS model in the fitting range $0.03 < k<0.2$ $\hompc$ for the monopole. However, at the end they combined their results with the configuration space result from Tamone et al. \cite{tamone_completed_2020}, which used the CLPT+GS RSD model with $s_{\rm min} = 32 \mpcoh$ or equivalently, $k_{\rm max} \sim 0.11 \hompc$.
Statistical only errors on $f\sigma_8^{\rm mg. \, \alpha s}$, $D_H$, and $D_M$ as calculated by de Mattia et al., are 23.4\%, 9.0\%, and 3.9\% 
, whereas Tamone et al. obtained 23.9\%, 9.1\%, and 4.6\%, respectively.
Combining these two measurements we find statistical fractional errors of 23.6\%, 9.0\% and 4.3\% on $f\sigma_8^{\rm mg. \, \alpha s}$, $D_H$, and $D_M$. Since the ELG likelihood in the BOSS DR16 sample cannot be adequately approximated with a Gaussian, instead of using the covariance matrix to fix the dilation parameters, we use the supplied probability grid in $D_M$ and  $D_H$, measuring the error on $f\sigma_8$ when $D_M$ and $D_H$
are constrained within a small range around their best-fit values. By doing so, we find a 15.2\% error on $f\sigma_8^{\rm fx. \, \alpha s}$. De Mattia et al. tested their analysis using different mock catalogues and found that the baseline Global Integral Constraint model (baseline, GIC) gives an 11.5\% error on $f\sigma_8^{\rm fx. \, \alpha s}$, which is shown in Figure~\ref{fig:Errors_vs_kmax}.

\subsubsection{Quasars}\label{subsec:QSO}
Quasar samples are of significant interest when it comes to constraining parameters at high redshift. In this study, we include the eBOSS DR16 sample containing $343,708$ Quasars covering $4,699$ deg$^2$ in the redshift range $0.8<z<2.2$, with an effective redshift of $z_{\rm eff} = 1.48$ (Neveux et al. \cite{neveux_completed_2020}). According to Croom et al. \cite{croom_2df_2005}, the quasar bias can be approximated as $b(z)_{\rm QSO} = 0.53 + 0.29 (1+z)^2$, which gives an effective bias of 2.31. Taking the average of best fits on $b_{\rm 1,NGC}\sigma_8$ and $b_{\rm 1,SGC}\sigma_8$ from their Table~10, and dividing by $\sigma_8(z_{\rm eff} = 1.48) = 0.41$, implies that the best fit effective bias is 2.33, which is very similar to the Croom formula. Note that BAO reconstruction is not applied to quasars, as their density is too low to provide us with an adequate estimate of the matter density field.

BAO-only measurements by Neveux et al. \cite{neveux_completed_2020} in Fourier space, combined with those from configuration space by Hou et al. \cite{hou_completed_2020}, gives the consensus errors of 2.6\%, 4.1\%, and 1.6\% for $D_M$, $D_H$, and $D_V$ (Table~6 in \cite{neveux_completed_2020}). By subtracting the systematic errors in quadrature, according to their Table~5, we find that these numbers change only slightly, and become 2.5\%, 4.0\%, and 1.5\% for $D_M$, $D_H$, and $D_V$.

\begin{figure}[t]
    \centering
    \includegraphics[width=1.0\textwidth,angle=0]{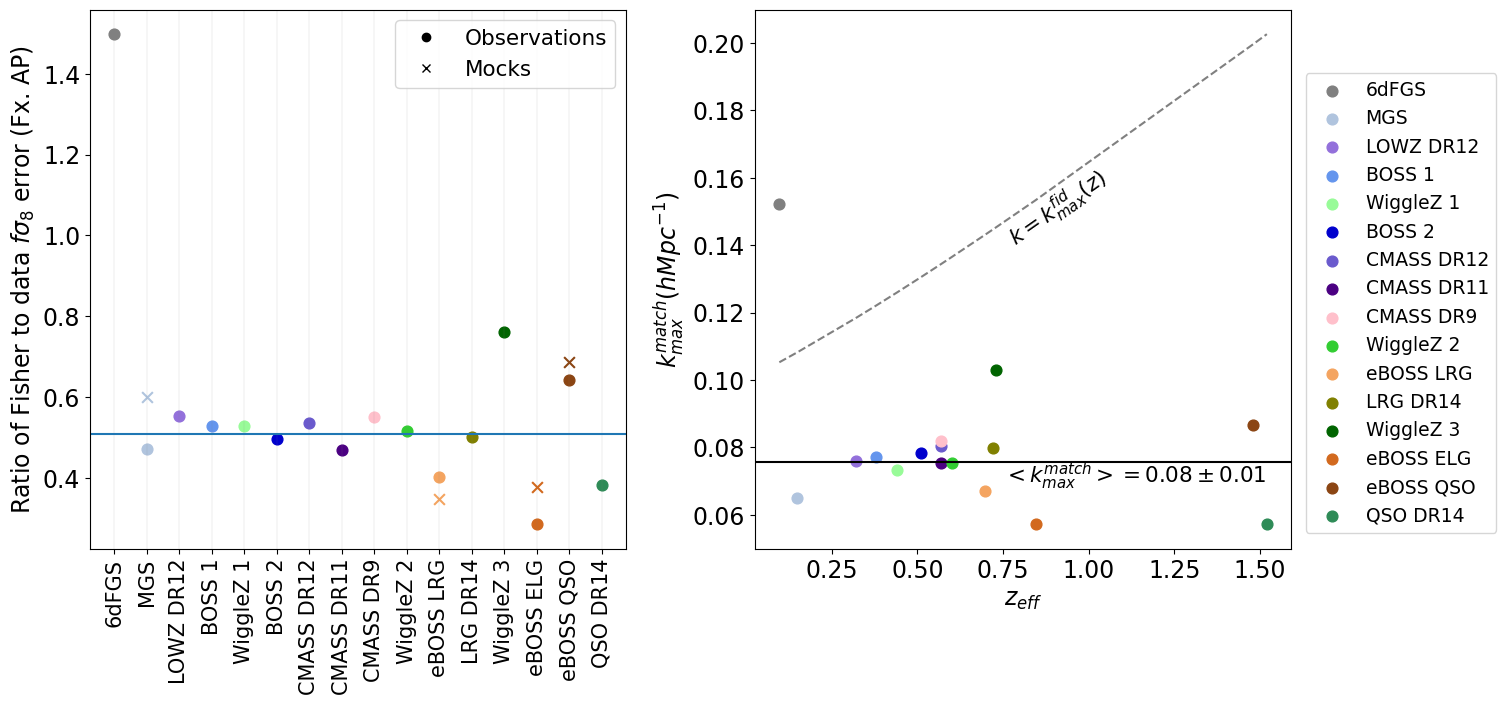}
    \caption{
    \emph{Left}: The difference between Fisher-based prediction $f \sigma_8$ error and observational $f\sigma_8^{\rm fx. \, \alpha s}$ error is plotted for each survey. The x markers indicate mock errors instead of the errors obtained from observations. 
    \emph{Right}: We have adjusted the $k_{\rm max}$ in order to make Fisher predictions match with the experiments. The  dashed line represents the fiducial value of $k_{\rm max} = 0.1 / D(z) \hompc$. Note that the fiducial cosmology used to draw this line is the same as the BOSS cosmology. (WiggleZ Near, Mid, and Far redshift slices are shortened as 1, 2, and 3)
    }
    \label{fig:kmax_vs_zeff}
\end{figure}

Neveux et al. \cite{neveux_completed_2020} used the TNS model for RSD measurements within the range $0.02<k<0.3$ $\hompc$, and obtained a 9.9\% error on $f\sigma_8^{\rm mg. \, \alpha s}$. Whereas Hou et al. \cite{hou_completed_2020} introduced a new method similar to TNS in order to model the correlation function within the scale range of $20<s_{\rm{min}}<160\mpcoh$) ($k_{\rm max} \sim 0.18$), and found a 10.9\% error on $f\sigma_8^{\rm mg. \, \alpha s}$. When combining the two to find a consensus result, this reduced to 9.7\%. Considering that both of these papers estimated that the systematic error is about 30\% of the statistical error, that gives a final only-statistical error of 9.3\% on $f\sigma_8^{\rm mg. \, \alpha s}$. 
After fixing the dilation parameters using the covariance matrix for the full shape analysis, and subtracting the 30\% systematic error in quadrature, we found that $f\sigma_8^{\rm fx. \, \alpha s}$ should have an error of 6.3\%.
Neveux et al. made use of 1000 EZmocks to estimate the covariance matrix and the observational systematic errors that we reported above. From these mocks catalogues, they recovered $f\sigma_8 = 0.467$, with an average fractional error of $0.059$, which is shown with a brown cross in Figure~\ref{fig:kmax_vs_zeff}.

In addition, we also run our Fisher-code for the eBOSS DR14 Quasar sample, consisting of $148,659$ galaxies with redshifts $0.8 < z < 2.2$ and an effective redshift of $z_{\rm eff} = 1.52$, and a footprint of $\rm 2,113 deg^2$. The BAO measurement for this sample showed a 3.8\% fractional error on $D_V$, considering that the systematic errors are negligible for this sample (Ata et al. \cite{ata_clustering_2018}).
The RSD measurement fitting the CLPT+GS model for scales larger than $s_{\rm min} = 20 \mpcoh$ ($k_{\rm max}\sim0.18\hompc$) showed a 16.4\% fractional statistical error on $f\sigma_8^{\rm mg. \, \alpha s}$. After taking systematic effects into account, the combined error rose to 18.5\%. From the covariance matrix provided, this translates to a 15.4\% error on $f\sigma_8^{\rm fx. \, \alpha s}$ (Table~9 of Zarrouk et al. \cite{zarrouk_clustering_2018}, stat. error). 
Zarrouk et al. also fitted for the galaxy bias and found that $b\sigma_8 = 1.038$, or $b(z_{\rm eff} = 1.52) = 2.6$. In our Fisher code, we re-scaled the Croom formula by a factor of 1.1 to match this bias.

\subsection{WiggleZ Dark Energy Survey}

The WiggleZ Dark Energy Survey was undertaken using the 3.9-meter Anglo-Australian Telescope (AAT; Drinkwater et al. \cite{drinkwater_wigglez_2010}). This survey observed approximately $240,000$ ELGs over the redshift range $0.2 < z < 1.0$ from 2006 to 2011. We consider the final data release that covers 816 $\rm deg^2$ in the sky in six regions, and contains $158,741$ ELGs. Following the analyses of Blake et al. \cite{blake_wigglez_2012} and Kazin et al. \cite{kazin_wigglez_2014}, we first combine these six regions together and divide the galaxies into three partially overlapping redshift bins $0.2<z_{\rm near}<0.6$, $0.4<z_{\rm mid}<0.8$, and $0.6<z_{\rm far}<1.0$, and then analyze each catalogue separately. Throughout our Fisher analysis, we assume that the galaxy bias is $b_{\rm ELG} = 0.8/D(z)$, to match the effective biases as assumed when performing reconstruction as in Kazin et al., in each redshift bin ($b_{\rm eff} =$ 1.0, 1.1, and 1.2 in Near, Mid, and Far, respectively)

Kazin et al. \cite{kazin_wigglez_2014}, found that after applying the reconstruction method by Padmanabhan et al., $D_V$ has an uncertainty of 4.8\%, 4.5\% and 3.4\% at effective redshifts $z_{\rm eff}^{\rm Near} = 0.44$, $z_{\rm eff}^{\rm Mid} = 0.60$, and $z_{\rm eff}^{\rm Far} = 0.730$, respectively.
Additionally, Blake et al. \cite{blake_wigglez_2012}, fitted with the RSD model introduced by Jennings et al. \cite{jennings_modelling_2010} (J+11), and for $k < 0.2 \hompc$ found a fractional error on $f\sigma_8^{\rm mg. \, \alpha s}$ of 19.4\%, 16.1\%, and 16.4\% for the near, middle, and far redshift slices, respectively. When fixing the dilation parameters, the fractional error bars reduce to $f\sigma_8^{\rm fx. \, \alpha s} = $ 13.2\%, 10.9\%, and 8.7\%.


\begin{figure}[t]
    \centering
    \includegraphics[width=0.7\textwidth,angle=0]{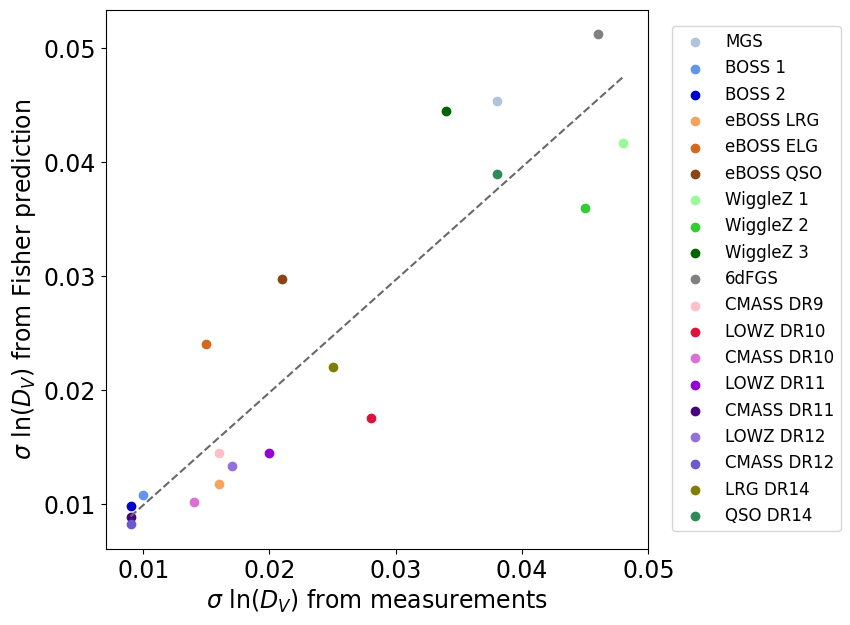}
    \caption{
    The Fisher-based predicted fractional error on $D_V$ plotted against the fractional error recovered from BAO-only analysis. The slope of the fitted line is $1.1$, which implies that the Fisher prediction and the BAO measurements are in good agreement. 
    }
    \label{fig:DV_obs_vs_Fisher}
\end{figure}

\section{Results}\label{sec:results}

We have performed a Fisher matrix analysis to predict BAO and RSD errors for the surveys introduced in Section~\ref{sec:Data}, matching the parameters assumed for the Fisher calculations as closely as we can to the parameters of the surveys. The final results of our calculations on the fractional error of the cosmological parameters, $D_M$, $D_H$, $D_V$, and $f\sigma_8$, for each survey's final data release and for BOSS and eBOSS intermediate data releases, are presented in Table~\ref{tab:results} and Table~\ref{tab:results2}, respectively (abbreviated as F.).
For comparison, we also present the errors of $D_M$, $D_H$, $D_V$ from the BAO measurements published in the most recent analyses of the data, with errors calculated using sets of mock catalogues (abbreviated as O. for observed).  In addition, we present the fractional errors of $f \sigma_8$ from the most recent RSD analysis of the data, when dilation parameters are marginalized over (abbreviated as Mg.\ $\alpha s$), and when they were held fixed (abbreviated as Fx.\ $\alpha s$). 
The RSD predictions depend strongly on the value of $k_{\rm max}$ adopted, which is the upper limit of the integration in Eq.~\ref{eq:Fisher2}. In the penultimate rows in Table~\ref{tab:results} and~\ref{tab:results2}, we present $k_{\rm max}^{\rm fid}=0.1/D(z_{\rm eff})$, matching the assumptions previously made for many surveys (e.g. \cite{zhao_extended_2016}), and $k_{\rm max}^{\rm match}$, that we will describe in detail later in this section.

In these tables, the effective redshift, $z_{\rm eff}$, is the weighted mean redshift, using weights $w_{\rm FKP} = (1+\bar{n}P_0)^{-1}$ from Feldman et al. \cite{feldman_power_1994}. 
The effective area, $A_{\rm eff}$, is the survey's area as reported from its corresponding clustering catalogue documentation. The effective volume is calculated from the equation below,
\begin{equation}
    V_{\rm eff} = V_{\rm survey} \sum_{i = 1}^{N_{Z}} \left(\frac{n(z_i)P_0}{1+n(z_i)P_0}\right)^2\,,
\end{equation}
where $P_0$ is the amplitude of the power spectrum at $k\sim0.15 \hompc$. Its value for 6dFGS, MGS, BOSS DR12 and eBOSS LRGs is $10,000$, for eBOSS ELGs is $4,000$, for eBOSS Quasars is $6,000$, for BOSS DR9-12 is $20,000$ and for WiggleZ is $5,000$, all in units of $h^{-3}$ Mpc$^3$.

For the combined 6dFGS and MGS samples, we obtain a 3.1\% accuracy on $D_V$, very similar to the 3.2\% error reported by Carter et al. \cite{carter_low_2018} for the combined samples. In Table~\ref{tab:results}, only the constraints recovered from the 6dFGS sample are reported, since we are interested in the surveys separately. Note that 6dFGS and MGS do not geometrically overlap more than 3\%, allowing their Fisher matrices to be added directly to obtain the final constraint at low redshift. Additionally, Alam et al. used the combined CMASS+LOWZ BOSS catalogue over the redshift range $0.2 < z < 0.75$, split into three partially overlapping catalogues in redshift covering $0.2 < z < 0.5$,  $0.4 < z < 0.6$, and $0.5 < z < 0.75$ and measured the BAO and RSD parameters in each of these bins separately. In this paper we did not analyze the high redshift bin separately since we combined the high redshift BOSS LRGs with the eBOSS LRGs to match the catalogue used by Bautista et al. These two samples geometrically overlap in the eBOSS footprint, and that is why we needed to calculate the overlapping area and the non-overlapping area separately and add the Fisher matrices to obtain the errors reported in Table~\ref{tab:results}.



\begin{center}

\begin{table}[t]
\caption{\label{tab:results} The Fisher forecast (shortened as F.) errors for $D_M$, $D_H$, $D_V$, and $f\sigma_8$ compared to the errors from the BAO and RSD measurements (shortened as O.). The RSD rows show $f\sigma_8$ errors from observations with and without marginalisation over the dilation parameters (Mg. for marginalized and Fx. for fixed).
For 6dFGS, the sample used for RSD has slightly different
$b_{\rm eff}$ and $z_{\rm eff}$ from the sample used for BAO;
in the Fisher forecast we use $b_{\rm eff}$ and $z_{\rm eff}$ from ref.~\cite{carter_low_2018} for both, for consistency.
The BOSS analysis have used 4 different RSD models that were briefly described in the text. The eBOSS $\rm LRG^{+}$ sample reported in this table is the combined eBOSS LRG sample covering $4,242 \rm \ deg^2$ with the BOSS DR12 sample at high redshift ($0.6<z<1.0$), covering $9,494 \rm \ deg^2$.
We also display the $k_{\rm max}$ of the observations (converted from configuration to Fourier space if applicable);
the fiducial Fisher $k_{\rm max}$; the $k_{\rm max}$ at which the Fisher RSD constraints
match the observations; the RSD model
used; whether the fit was performed
in configuration ($s$) or Fourier ($k$)
space; and the references for the RSD and BAO papers.
\\ }
\begin{tabular}{||c||c|c|c c|c c c|c c c||}
\hline\hline
  & 6dFGS
  & MGS 
  & \multicolumn{2}{c|}{BOSS} 
  & \multicolumn{3}{c|}{eBOSS} & \multicolumn{3}{c||}{WiggleZ}\\
  &
  & 
  & Near  &  Mid
  & $\rm LRG^{+}$ & ELG & QSO 
  & Near&	Mid&	Far
  \\
\hline
$z_{\rm eff}$ 
& 0.097 & 0.15 & 0.38 & 0.51 & 0.698 & 0.845 & 1.48 & 0.44 & 0.600 & 0.73 \\

$A_{\rm eff}(deg^2)$  
& 17000 & 6813 & 9329 & 9329 & 4242 & 727 & 9494 & 816 & 816 & 816\\

$V_{eff}(Gpc^3)$ & 0.134& 0.282 & 3.401 & 4.082 & 2.652 & 0.522 & 0.525 & 0.288 & 0.364 & 0.223\\

$b_{\rm eff}$
&1.65&1.5&2.03&2.13&2.2&1.52&2.33&1.0&1.1&1.2\\

\hline\hline
\multicolumn{11}{c}{Fractional errors from the BAO analysis (in percent)}\\
\hline\hline

$D_M$ (O.)&
&  &  1.5&	1.4&    1.6&       4.3& 	2.5&		 &     &  \\

$D_M$ (F.)&
8.0&  6.4&	1.5&	1.3&	1.7&	3.6&	3.4&	6.9&	6.0&	7.5\\	
\hline

$D_H$ (O.)&
& 	 &  2.7&	2.3&	2.5&	  9.0&	4.0&		 &		 &       \\
$D_H$ (F.)&
16.1&  13.8&	3.1&	2.8&	3.2&	6.3&	5.0&	11.3&	9.5&	11.4\\
\hline

$D_V$ (O.)&
4.6&    3.8&    1.0&	0.9&    1.6&    2.7&  1.5&	4.8&	4.5&	3.4	\\
$D_V$ (F.)&
5.8&    4.9&    1.1&	1.0&	1.2&	2.5&	3.0&	4.6&	3.9&	4.9	\\

\hline\hline

BAO ref. 
& \cite{carter_low_2018}
& \cite{ross_clustering_2015}
&\cite{alam_clustering_2017}
&\cite{alam_clustering_2017}
&\cite{bautista_completed_2020}
&\cite{de_mattia_completed_2020}
&\cite{neveux_completed_2020}
&\cite{kazin_wigglez_2014}
&\cite{kazin_wigglez_2014}
&\cite{kazin_wigglez_2014}\\

\hline\hline
\multicolumn{11}{c}{Fractional errors from the RSD analysis (in percent)}\\
\hline\hline

$f \sigma_8$(O.Mg.)&
&  40.5&	7.8&	7.6&	7.9&
23.6&	9.3&	19.4&	16.1&	16.4	\\
$f \sigma_8$(O.Fx.)&
13.0&  31.8&	7.0&	6.4&	7.6&	 15.2&  6.3&	13.2&	10.9&	8.7	    \\
$f \sigma_8$(F.Fx.)&
19.5&   15.0&	3.7&	3.2&	3.1&	4.4&	4.0&	7.0&	5.6&	6.6\\

\hline\hline
\multicolumn{11}{c}{$k_{max}$ for the RSD modeling}\\
\hline\hline

$k_{\rm max}$(O.)&
0.23&  0.14&  0.17&  0.17&  0.16&
0.16&	0.24&	0.2&	0.2&	0.2\\

$k_{\rm max}$(Fid.\, F.)&
0.11&  0.11&	0.12&	0.13&	0.14&	0.15&	0.20&	0.12&	0.13&	0.14	\\
$k_{\rm max}$(Match)&
0.16&   0.06&	0.08&	0.08&	0.07&	0.06&	0.08&	0.07&	0.07&	0.11 \\

\hline
RSD Model&
Sc.  &  CLPT&  
\multicolumn{2}{c|}{4 Models}& 
\multicolumn{3}{c|}{TNS\&CLPT+GS}&
\multicolumn{3}{c||}{J+11}

\\

 RSD space&
 $s$ &
 $s$ &
 $k$/$s$ &
 $k$/$s$ &
 $s$ &
 $k$/$s$ &
 $k$/$s$ &
 $k$ &
 $k$ &
 $k$ \\

RSD ref. 
& \cite{beutler_6df_2012}
& \cite{howlett_clustering_2015}
&\cite{alam_clustering_2017}
& \cite{alam_clustering_2017}
&\cite{bautista_completed_2020}
&\cite{de_mattia_completed_2020}
&\cite{neveux_completed_2020}
&\cite{blake_wigglez_2012}
&\cite{blake_wigglez_2012}
&\cite{blake_wigglez_2012} \\

\hline\hline
\end{tabular}

\end{table}

\end{center}

In the following, we briefly describe the key results that our analysis implies.
\begin{itemize}

    \item From the left panel in Figure~\ref{fig:Errors_vs_Veff}, we are able to confirm that there is a linear relationship between the fractional error on $D_V$ and $V_{\rm eff}^{-1/2}$, as expected from Eq.~\ref{eq:SE}, both for observations and predictions. The Fisher-based $D_V$ errors were calculated under the assumption of $k\gtrsim0.5 \hompc$. From the left panel in Figure~\ref{fig:Errors_vs_kmax}, we found that $\sigma_{\ln{D_V}}$ is not sensitive to the choice of $k_{\rm max}$ for $k\gtrsim0.3 \hompc$ (Section~\ref{subsec:kmax} for more detail on the $k_{\rm max}$ choice). 
    Furthermore, we obtained similar slopes for the Fisher analysis and observations, showing that the Fisher method works well, and that observational methods are able to extract information reaching close to the maximum possible. In addition,
    Figure~\ref{fig:DV_obs_vs_Fisher} supports this claim by revealing a linear relationship, with a slope very close to unity, between the measured $D_V$ error and the Fisher-predicted $D_V$ error.

    \item Table~\ref{tab:results} and~\ref{tab:results2} show that our Fisher prediction of fractional error on the volume averaged distance from the BAO measurements, $\sigma_{D_V} / D_V$, deviates less than 35\% from  analyses that use similar techniques, including reconstruction and covariance matrices calculated using mock catalogues. 
    We consider this to be an exceptionally positive result for the field of BAO analyses, showing that current methods are extracting all of the available signal (i.e.\ reconstruction is working as well as expected in Seo \& Eisenstein, with a 50\% drop in $\Sigma_s$). Better reconstruction methods (e.g. \cite{birkin_reconstructing_2019, sarpa_bao_2019, modi_cosmological_2018}) would be able to improve the BAO constraints by reducing the non-linear damping.

    \item The picture is worse for RSD-based measurements: If we assume that $k_{\rm max} = k_{\rm max}^{\rm fid}$, the $f\sigma_8$ fractional error is not very well estimated by the Fisher matrix analysis, and it shows a lower correlation with the effective volume than $D_V$ (Figure~\ref{fig:Errors_vs_Veff}; right panel). In our baseline Fisher calculation, we assume that the dilation parameters are fixed. Therefore, we choose the $f\sigma_8$ errors taken from observations to be those for which the dilation parameters $D_M$ and $D_H$ were fixed. We have been able to adjust most of the measurements we have considered to match the baseline assumed for the Fisher predictions. Whether or not we marginalise has a big effect on the results, and so it is important to match Fisher with observations.

    \item The $f\sigma_8$ error is highly dependent on the choice of maximum $k$ to which we integrate to make Fisher predictions and hence from which we can recover information. Decreasing $k_{\rm max}$ weakens the predicted constraints on $f\sigma_8$ (Figure~\ref{fig:Errors_vs_kmax}; right panel). For most of the catalogues used in this work, we calculated the $k_{\rm max}^{\rm match}$ that, when used in the Fisher analysis, gives an $f\sigma_8$ error that matches with that of the experiments. The average of $k_{\rm max}^{\rm match}$ over these catalogues is about $0.08 \pm 0.01 \hompc$, with remarkably little scatter around this value. This should be interpreted as the amount of linear information that can be recovered from these data.

    If the theoretical arguments that led to expecting $k_{\rm max}^\mathrm{fid}$ to increase at higher redshifts hold true, then the data analyses are missing significant information, particularly at high redshift. Alternatively, the theoretical arguments are wrong, possibly because the high bias of the samples analysed at high redshift shifts the non-linear scale to lower $k$. Either way, this is clearly an area where improved modeling is required (e.g.
    \cite{carrasco_effective_2012,baumann_cosmological_2012,ivanov_constraining_2020,damico_cosmological_2020,chen_redshift-space_2021,
    kokron_cosmology_2021,zhai_aemulus_2019}).
    
    \item 
    The right and left panels of Figure~\ref{fig:kmax_vs_zeff} show a consistent story. The left panel shows that the ratio of Fisher-predicted to observed $f \sigma_8$ error generally decreases to higher redshift (note that the samples are arranged from low to high redshift). The right hand panel shows that to match the observed information, we need $k_{\rm max}^\mathrm{match}$ to stay fixed, while $k_{\rm max}^\mathrm{fid}$ increases with redshift. The quasars are least consistent with this trend, with a relatively better
    agreement between Fisher and observations, but a strong change in $k_{\rm max}$. Figure~\ref{fig:Errors_vs_kmax} explains why this is: it shows that, for the low galaxy density of the quasar sample, there is increasingly little information on small scales due to the high shot noise. A large value of $k_{\rm max}^\mathrm{fid}$ brings in less expected information (the green and brown curves are flatter, showing that for eBOSS DR14 and DR16 QSO information saturates quickly). 
    This makes it even more interesting that, for all samples, the information recovered on $f \sigma_8$ is that predicted by the Fisher analysis to a consistent $k_{\rm max}^{\rm match}$,
    despite large differences in shot noise, bias, and redshift.

    \item Our baseline results are presented assuming that the dilation parameters are fixed. We have also considered the alternative of marginalizing over unknown dilation parameters, using the sample itself, predominantly through the BAO signal to constrain these. We compare this alternative procedure for both the data and the Fisher matrix. In general, we find a better agreement between the Fisher and data for this procedure: comparing with the case of fixed dilation parameters shows that fixing the $\alpha$s does not constrain $f\sigma_8$ as much in the data as in the Fisher. We suggest that the reason for this is that when modeling the data we have other parameters fitted (e.g.\ shot noise, nonlinear bias parameters), and so our $f\sigma_8$ errors assuming fixed dilation do not reach their theoretical minima. That is, for the Fisher calculation, we reach a precision when we fix the dilation parameters that goes beyond that achievable in the data for other reasons.
    
    Marginalising over the dilation parameters leads to more scatter in the ratio of the Fisher-based error on $f \sigma_8^{\rm mg.\,\alpha s}$ to the data error. We also find a small shift in the value of $k_{\rm max}^{\rm match}$ required to match Fisher and data: marginalising over the dilation parameters, we find $k_{\rm max}^{\rm match}=0.09 \pm 0.02 \hompc$. The increased scatter leads to an error on $k_{\rm max}^{\rm match}$ of $0.02\hompc$ which is twice as large for fixed dilation parameters. This can be seen by eye comparing Figure~\ref{fig:kmax_vs_zeff_marg} to Figure~\ref{fig:kmax_vs_zeff}. This is likely because in the data,
    imperfect measurement of the BAO peak adds an extra
    source of scatter in the marginalized $f\sigma_8$ errors.
    Hence our fiducial comparison uses $f\sigma_8$ errors with fixed dilation parameters, which are thus less noisy.
    In summary, our conclusion remains unchanged when considering $f \sigma_8^{\rm mg.\,\alpha s}$; we still do not see the error decrease to high redshift, as one would naively expect. The exact prescription we use for marginalising over the dilation parameters is presented in Appendix~\ref{appendix_A}.
    
    \item By considering the average measurement recovered from mock catalogues, we are able to remove the statistical error from the measurements: i.e.\ the actual observation is only one realization of what could have happened in that particular region of the universe, and therefore its constraints can depend on how lucky the experiment was. For the MGS, eBOSS LRG, eBOSS ELG, and eBOSS Quasar experiments, the teams created 1000 mock catalogues and released the results of fitting the models to them, finding an $f \sigma_8$ error of 25.0\%, 8.8\%, 11.5\%, and 5.9\%, respectively. We plotted the ratio of Fisher forecast to the $f \sigma_8$ error recovered from both observations and mocks, if available, in the left panel of Figure~\ref{fig:kmax_vs_zeff}.  We see that the trends seen between Fisher predictions and measurements are well matched between data and mocks. This suggests that the trends are due to the methodology (particularly the model used to fit to the data) rather than statistical fluctuations.

\end{itemize}

\begin{center}

\begin{table}[t]
\centering
\caption{\label{tab:results2} Same as Table \ref{tab:results}, but including intermediate data releases of BOSS and eBOSS. LOWZ and CMASS are shortened as LZ and CM. \\}
\begin{tabular}{||c||c|c c|c c|c c|c c|c c||}
\hline\hline
  & BOSS & \multicolumn{2}{c|}{BOSS DR10} & \multicolumn{2}{c|}{BOSS DR11} &
  \multicolumn{2}{c|}{BOSS DR12} &
  \multicolumn{2}{c|}{eBOSS DR14} &
  \multicolumn{2}{c||}{eBOSS DR16}
  \\

  & DR9
  & LZ & CM 
  & LZ & CM 
  & LZ & CM 
  & $\rm LRG^{+}$  & QSO
  & $\rm LRG^{+}$  & QSO
  \\
  
\hline
$z_{\rm eff}$&
0.57&	0.32&	0.57& 	0.32&	0.57& 0.32&	0.57&	0.72&	1.52&  0.698&  1.48\\

$A_{\rm eff}(deg^2)$ & 
3275&	5156&	6161&	7341&	8377&  8337&	9376&	1845&	2113&  9494&   4699\\

$V_{eff}(Gpc^3)$&

2.474&	1.770&	4.669&	2.527&	6.381&  2.185&	5.034&	0.910&	0.218&  2.652&  0.525\\

$b_{\rm eff}$& 
2.0&  1.8&  2.11& 1.85&  2.05&   1.9&   2.1&   2.0&  2.63 & 2.2& 2.33\\

\hline\hline
\multicolumn{12}{c}{Fractional errors from the BAO analysis (in percent)}\\
\hline\hline

$D_M$(O.)&
2.0&	&	1.9&	&	1.4&  2.2&   1.3&	 &	 &  1.6&   2.5  \\

$D_M$(F.)&
2.2&	2.7&	1.6&	2.2&	1.4&	2.0&	1.3&	3.2&	5.2&    1.7&    3.4\\ 
\hline

$D_H$(O.)&
3.8&	&	5.0&	&	3.5& 5.9& 2.9&	&	 & 2.5&   4.0\\

$D_H$(F.)&
4.4&	5.4&	3.1&	4.5&	2.7&	4.2&	2.5&	5.9&	7.8&    3.2&    5.0\\
\hline

$D_V$(O.)&
1.6&	2.8&	1.4&	2.0&	0.9&  1.7&  0.9&  2.5&	3.8& 1.6&  1.5\\

$D_V$(F.)&
1.6&	2.0&	1.1&	1.6&	1.0&	1.5&	0.9&	2.2&	3.9&    1.2&    3.0\\
\hline

BAO ref. 
& \cite{anderson_clustering_2012} 
&\cite{anderson_clustering_2014}
&\cite{anderson_clustering_2014}
&\cite{anderson_clustering_2014}
&\cite{anderson_clustering_2014}
&\cite{gil-marin_clustering_2016-2}
&\cite{gil-marin_clustering_2016-2}
&\cite{bautista_sdss-iv_2018}
&\cite{ata_clustering_2018}
&\cite{bautista_completed_2020}
&\cite{neveux_completed_2020}\\

\hline\hline
\multicolumn{12}{c}{Fractional errors from the RSD analysis (in percent)}\\
\hline\hline

$f \sigma_8$(O.Mg.)&

14.6&   23.3&   12.8&   20.8&	9.9&   15.7&   8.6&	29.2&	16.4& 7.9& 9.3\\

$f \sigma_8$(O.Fx.)&
8.1&       &	    &      &	6.0&	9.1&   5.0&    10.4&	15.4& 7.6&  6.3\\

$f \sigma_8$(F.Fx.)&
4.5&	6.2&	3.3&	5.3&	2.8&	5.0&	2.7&	5.2&	5.9&    3.1&    4.0\\

\hline\hline
\multicolumn{12}{c}{$k_{max}$ for the RSD modeling}\\
\hline\hline

$k_{\rm max}$(O.)&
0.14& 0.09& 0.09& 0.09& 0.14&
0.17&	0.17&	0.13&	0.18&	0.16&	0.24\\

$k_{\rm max}$(Fid. F.)&
0.13&	0.12&	0.13&	0.12&	0.13&	0.12&	0.13&	0.15&	0.20&  0.14&  0.20\\
$k_{\rm max}$(Match)&
0.08&	&	&	&	0.07&	0.08&	0.08&	0.08&	0.07&  0.07& 0.08\\

\hline
RSD Model&
R+11
& & &
& R+11 
&
\multicolumn{2}{c|}{TNS}&
\multicolumn{2}{c|}{CLPT+GS}&
\multicolumn{2}{c||}{TNS\&CLPT}\\

 RSD space&
 $s$ &
 $s$ &
 $s$ &
 $s$ &
 $s$ &
 $k$ &
 $k$ &
 $s$ &
 $s$ &
 $s$ &
 $k$/$s$ \\

RSD ref. 
& \cite{reid_clustering_2012}
& \cite{sanchez_clustering_2014}
& \cite{sanchez_clustering_2014}
& \cite{sanchez_clustering_2014}
&\cite{samushia_clustering_2014}
&\cite{gil-marin_clustering_2016}
&\cite{gil-marin_clustering_2016}
&\cite{icaza-lizaola_clustering_2020}
&\cite{zarrouk_clustering_2018} &\cite{bautista_completed_2020}
&\cite{neveux_completed_2020}\\

\hline\hline

\end{tabular}
\end{table}
\end{center}

\section{Discussion}\label{sec:discussion}

We have performed a Fisher analysis for 19 different survey catalogues to obtain the constraints on the co-moving angular diameter distance, $D_M$, the Hubble distance, $D_H$, and the growth rate of structure $f\sigma_8$, from BAO and RSD analyses. Furthermore, the statistical errors derived were compared for each catalogue to that recovered in recent analyses modelling the correlation function (or the power spectrum). We only selected studies that adopted similar methodologies, using covariance matrices for the 2-point measurements created using mock catalogues, and for BAO measurements using a basic reconstruction method (except for the quasar samples where reconstruction does not improve the BAO signal).

Our Fisher-based BAO and RSD predictions are well matched to those from Zhao et al. \cite{zhao_extended_2016} for the eBOSS survey if we correct for the difference between predicted and actual survey parameters (area, galaxy redshift distribution, galaxy bias). The differences between predicted and actual survey details make less than a 60\% change on the measured errors. 
By comparing the final measurements with Fisher predictions using the actual eBOSS details, we find that the Fisher predictions and observational results are within 30\% of each other for all samples for errors on $D_V$. 
In contrast, for RSD we find that in general, the errors predicted from our Fisher analysis are about 50\% of those of the observations.

Using Fisher-based predictions for the BAO constraints on the BOSS Near sample ($0.2 < z < 0.5$), Dawson et al. \cite{dawson_baryon_2013} predicted a 1\% error on $D_M$ and a 1.8\% error on $D_H$, and from their results it can be inferred that the error on $D_V$ is about 1.0\%, which is in a good agreement with our Fisher-prediction for BOSS DR12 Near sample. After the BOSS DR9 was released, Font-Ribera et al. \cite{font-ribera_desi_2014} performed a Fisher matrix analysis and found 7.0\% fractional error on $f \sigma_8$ and 1.15\% fractional error on $D_V$ for DR9 CMASS. 

More accurate RSD techniques allow for extracting RSD information to smaller scales. For instance, Reid et al. \cite{reid_25_2014} were able to extend the RSD analysis to scales of $0.8 \mpcoh$ in configuration space, using an HOD to model the small-scale RSD, and found 2.4\% accuracy on $f \sigma_8$ in the BOSS CMASS DR10 sample. This is a factor of 1.4 smaller than our Fisher matrix prediction using $k_{\rm max}^{\rm fid} = 0.132$. 
For the Fisher-based analysis to be as small as this error, we would need to include linear information to $k_{\rm max}^{\rm match} = 0.19 \hompc$.
Additionally, using small scale RSD modelling, Lange et al. \cite{lange_five-percent_2021} extracted the RSD information down to scales of about $0.4 \mpcoh$ for two volume-limited catalogues created from the BOSS LOWZ sample in the North Galactic Cap. They found a precision of 5.1\% and 5.8\% on $f \sigma_8$ at effective redshifts of 0.25 and 0.4, respectively. Our Fisher code, when run on their catalogue with the fiducial $k_{\rm max}$ = 0.114 and 0.123 $\hompc$, yields an error of 11.8\% and 9.8\% on $f \sigma_8$. This implies that $k_{\rm max}^{\rm match}$ should be $0.23 \hompc$ and $0.14 \hompc$, respectively. 
This shows that improved small-scale modelling can extract information to smaller scales than the apparently-universal
$k_{\rm max} = 0.08 \hompc$ we found for the large-scale analyses. However, ensuring that the RSD modelling remains unbiased on very small scales remains a challenge. 

Our analyses suggest that Fisher matrix calculations are very good at predicting BAO constraints, and that measurements and the way in which they are analysed are delivering the expected level of precision. This is clearly great news for future projects. The same could be true for RSD if $k_{\rm max}$ is chosen correctly--our work suggests that current methods extract RSD information to a fixed $k_{\rm max}$ for all samples and redshifts. For both measurements, there is room for improvement--particularly with reconstruction for BAO, and fully extracting the linear signal from which to measure RSD. 

It is possible to interpret the consistent $k_{\rm max}$ as indicating that the models fitted do not allow for an increase in the linear information at higher redshift, rather than the information being missing. Indeed, recent analyses using emulators to extract information from smaller scales have been able to provide strong improvements on the precision obtained from RSD \cite{zhai_aemulus_2019, lange_five-percent_2021, DeRose21, kobayashi_accurate_2020}. The improvement from small-scale analyses shows that the information is available, and improved modelling may be able to
extract it.


\section*{Acknowledgments}
We would like to thank Mariana Vargas Magana and Richard Neveux for providing the covariance matrices for eBOSS Quasars, and Chris Blake and Hector Gil-Marín for thoughtful comments and useful discussions. 
Special acknowledgment to Yuting Wang who provided code and results for eBOSS Quasar DR14 for comparison to our results.
This research made use of the PYTHON packages NUMPY (Walt,
Colbert \& Varoquaux 2011; \cite{numpy}), SCIPY (Jones et al.; \cite{Scipy}), MATPLOTLIB
(Hunter 2007; \cite{Matplotlib}), COLOSSUS (Benedikt Diemer; 2018 \cite{diemer_colossus_2018}), and ASTROPY (Astropy Collaboration 2013; \cite{the_astropy_collaboration_astropy_2013}). This research was enabled in part by support provided by Compute Ontario (www.computeontario.ca) and Compute Canada (www.computecanada.ca). Research at Perimeter Institute is supported in part by the Government of Canada through the Department of Innovation, Science and Economic Development Canada and by the Province of Ontario through the Ministry of Colleges and Universities. 

\appendix
\section{Marginalizing over dilation parameters}\label{appendix_A}

Our baseline, throughout this paper, when analyzing the $f\sigma_8$ constraint with RSD, is to fix the dilation parameters. In this section, we show that our conclusions remain unchanged if we instead marginalize over the dilation parameters. Considering the free parameters in the Fisher matrix in Eq.~\ref{eq:Fisher} to be $\{\alpha_\perp, \alpha_\parallel, \ln{f\sigma_8}, \ln{b\sigma_8}\}$, we can build a 4 by 4 Fisher matrix using the following derivatives (\cite{samushia_effects_2010}):

\begin{equation}\label{eq:partial_RSD_4by4}
    \begin{aligned}
        \frac{\partial \ln{P}}{\partial \alpha_{\perp}} = 
        -2
        +4f\sigma_8(z)\mu^2(1-\mu^2)/(b\sigma_8(z) + f\sigma_8(z) \mu^2)
        -(1-\mu^2)\frac{\partial \ln{P}}{\partial\ln(k)}
        \,, \\
        \frac{\partial \ln{P}}{\partial \alpha_{\parallel}} = 
        -1
        -4f\sigma_8(z)\mu^2(1-\mu^2)/(b\sigma_8(z) + f\sigma_8(z) \mu^2)
        -\mu^2\frac{\partial \ln{P}}{\partial\ln(k)}
        \,,\\
        \frac{\partial \ln{P}}{\partial \ln{f\sigma_8}} = \frac{2\mu^2 f \sigma_8(z)}{b\sigma_8(z) + f\sigma_8(z) \mu^2} 
        \,,\\
        \frac{\partial \ln{P}}{\partial \ln{b\sigma_8}} = \frac{2b\sigma_8(z)}{b\sigma_8(z) + f\sigma_8(z) \mu^2}
        \,.
    \end{aligned}
\end{equation}

\begin{figure}[h]
    \centering
    \includegraphics[width=1.0\textwidth,angle=0]{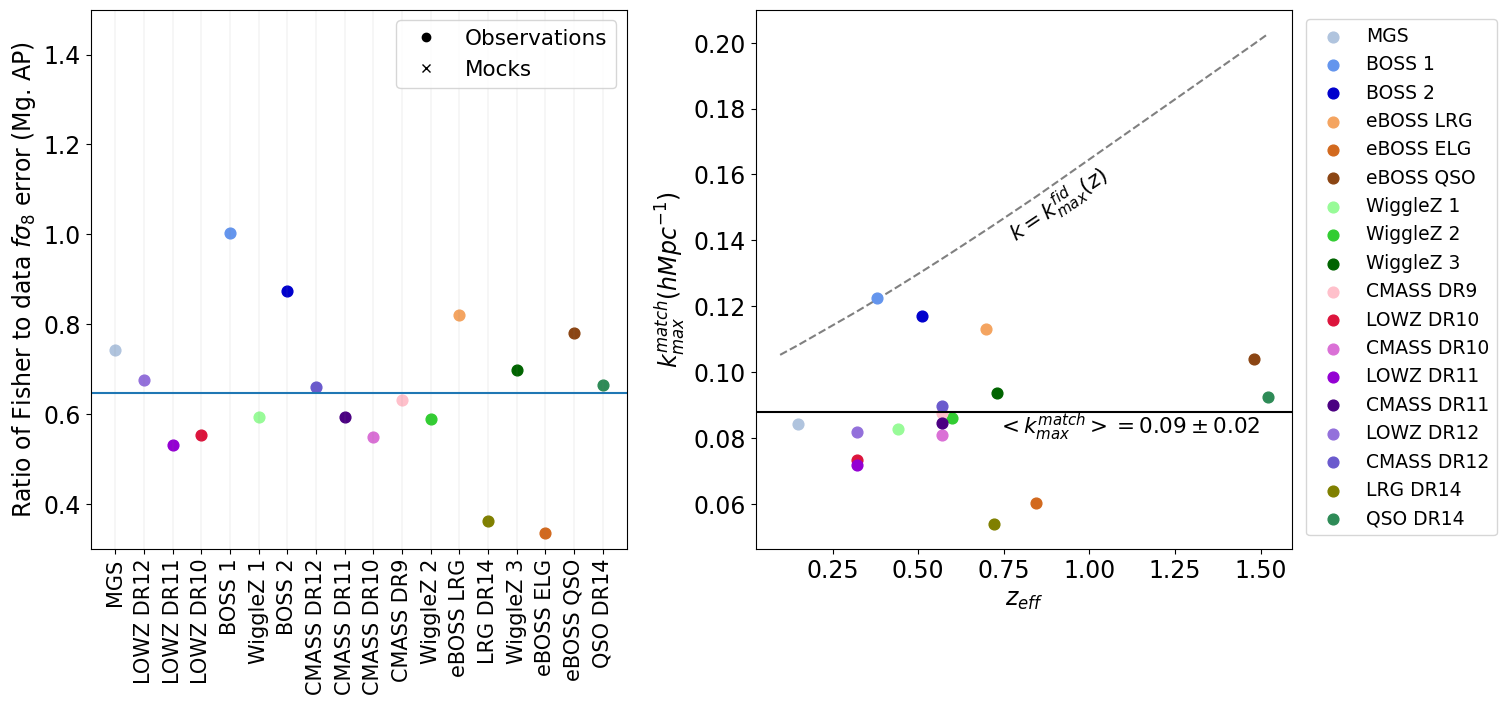}
    \caption{
    \emph{Left}: The difference between Fisher-based predicted $f \sigma_8$ error and observational $f\sigma_8^{\rm mg. \, \alpha s}$ error is plotted for each survey. 
    \emph{Right}: We have adjusted the $k_{\rm max}$ in order to make Fisher predictions match with the experiments. The dashed line represents the fiducial value of $k_{\rm max} = 0.1 / D(z) \hompc$. (Near, Mid, and Far redshift slices are shortened as 1, 2 and 3, respectively.)
    }
    \label{fig:kmax_vs_zeff_marg}
\end{figure}

Although we are fitting to the $\alpha s$, and relying on the BAO signal to constrain these, we follow the standard approach and assume that reconstruction cannot be used, because we also wish to model the form of the clustering signal - something that is hard to do post-reconstruction. Thus, our errors on $\alpha$ will be larger than those in Table~\ref{tab:results} and~\ref{tab:results2}, where a standard reconstruction technique was applied to both Fisher prediction and the data. Assuming $k_{\rm max}^{\rm fid} = 0.1/D(z) \hompc$, we predict errors for these free parameters. The fractional error for $f \sigma_8^{\rm mg.\,\alpha s}$ is reported in Table~\ref{tab:results3} and Table~\ref{tab:results4}. Figure~\ref{fig:kmax_vs_zeff_marg} is analogous to Figure~\ref{fig:kmax_vs_zeff}, except that it is calculated for $f\sigma_8^{\rm mg. \,\alpha s}$. As can be seen from this plot, the scatter around the Fisher to data ratio is larger compared to Figure~\ref{fig:kmax_vs_zeff}, and therefore, so is the scatter around $k_{\rm max}^{\rm match}$. 

\begin{figure}
    \centering
    \begin{subfigure}{.5\textwidth}
        \centering
        \includegraphics[width=1.0\textwidth,angle=0]{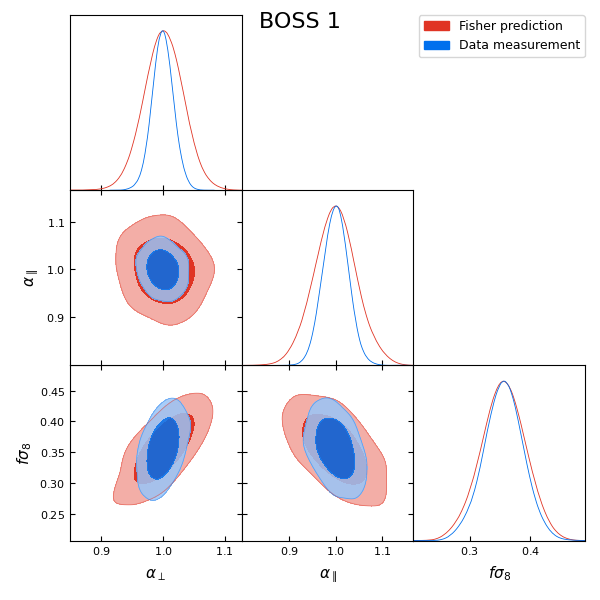}
    \end{subfigure}%
    ~
    \begin{subfigure}{.5\textwidth}
        \centering
        \includegraphics[width=1.0\textwidth,angle=0]{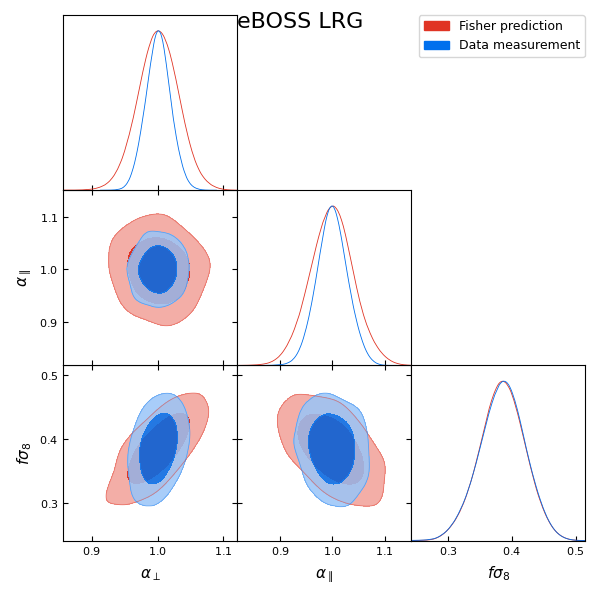}
    \end{subfigure}
    \caption{Constraints on $\alpha_{\perp}$, $\alpha_{\parallel}$, and $f\sigma_8$. Red contours show 68 and 95 percent confidence regions for the Fisher analysis with $k_{\rm max} = 0.09 \hompc$ (this work). \emph{Left:} The full-shape consensus analysis of the BOSS Near DR12 sample is shown in blue (Alam et al. \cite{alam_clustering_2017}). \emph{Right:} The full-shape RSD analysis for eBOSS LRG DR16 in configuration space analysis is shown in blue (Bautistia et al. \cite{bautista_completed_2020}).
    }
    \label{fig:BOSS_Near_eBOSS_LRG}
\end{figure}

\begin{figure}
    \centering
    \begin{subfigure}{.5\textwidth}
    \centering
    \includegraphics[width=1.0\textwidth,angle=0]{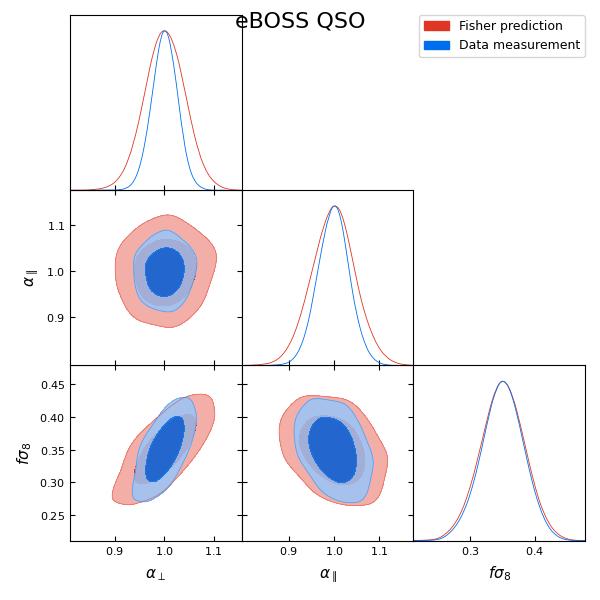}
    \end{subfigure}%
    ~
    \begin{subfigure}{.5\textwidth}
    \centering
    \includegraphics[width=1.0\textwidth,angle=0]{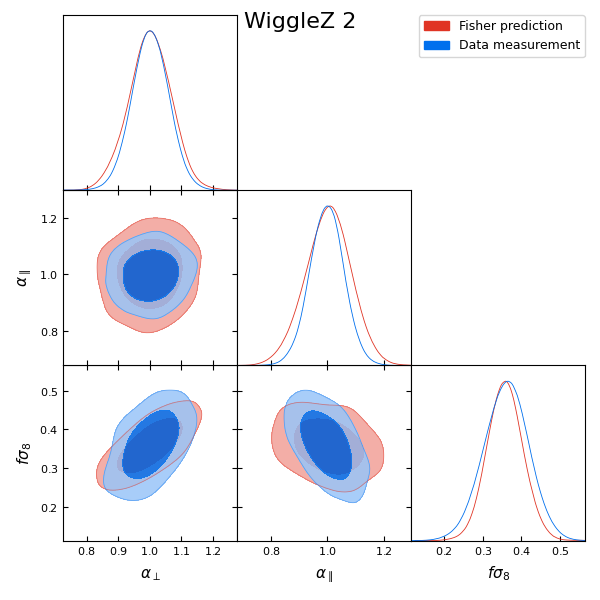}
    \end{subfigure}
    \caption{Constraints on $\alpha_{\perp}$, $\alpha_{\parallel}$, and $f\sigma_8$. Red contours show 68 and 95 percent confidence regions for the Fisher analysis with $k_{\rm max} = 0.09 \hompc$ (this work). \emph{Left:} The full-shape RSD analysis of the eBOSS Quasar DR16 sample is shown in blue (Neveux et al. \cite{neveux_completed_2020}).
    \emph{Right:} The joint fit to expansion and growth for WiggleZ Mid redshift slice is shown in blue. The covariance matrix was originally in $\{A,F,f\sigma_8\}$ but then we converted to $\alpha_{\parallel}$, $\alpha_{\perp}$ and $f\sigma_8$ (Blake et al. \cite{blake_wigglez_2012}).}
    \label{fig:eBOSS_Quasar_DR16_WiggleZ}
\end{figure}

\begin{figure}
    \centering
    \includegraphics[width=0.5\textwidth,angle=0]{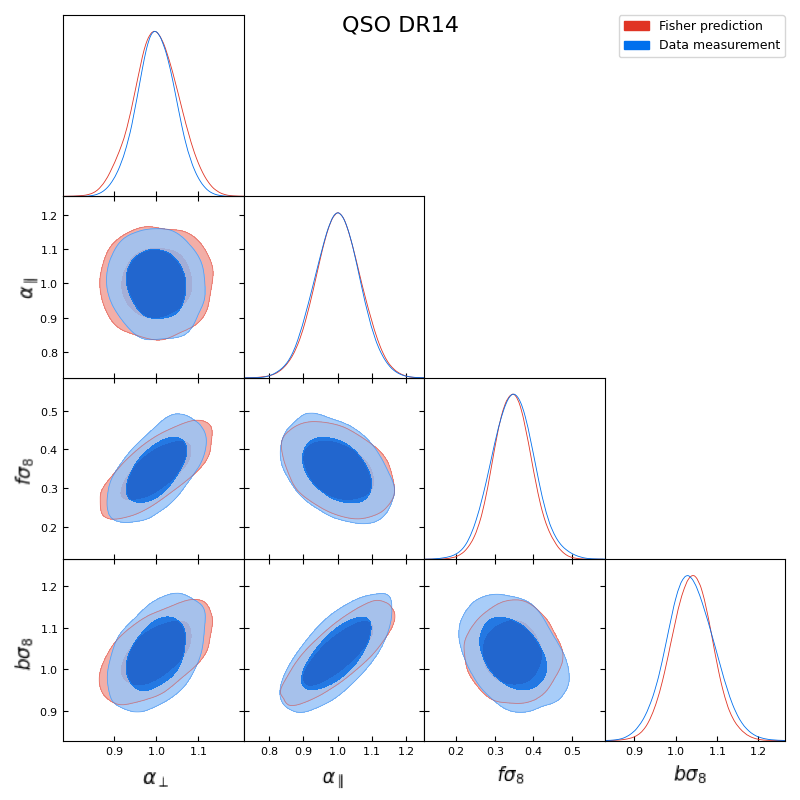}
    \caption{Constraints on $\alpha_{\perp}$, $\alpha_{\parallel}$, $f\sigma_8$, and $b\sigma_8$. Red contours show 68 and 95 percent confidence regions for the Fisher analysis with $k_{\rm max} = 0.09 \hompc$ (this work), and the blue contours show confidence regions for eBOSS Quasar DR14 sample analysis from 5-parameter RSD modeling by Zarrouk et al. \cite{zarrouk_clustering_2018}.}
    \label{fig:eBOSS_Quasar_DR14}
\end{figure}

\begin{figure}
    \centering
    \includegraphics[width=0.5\textwidth,angle=0]{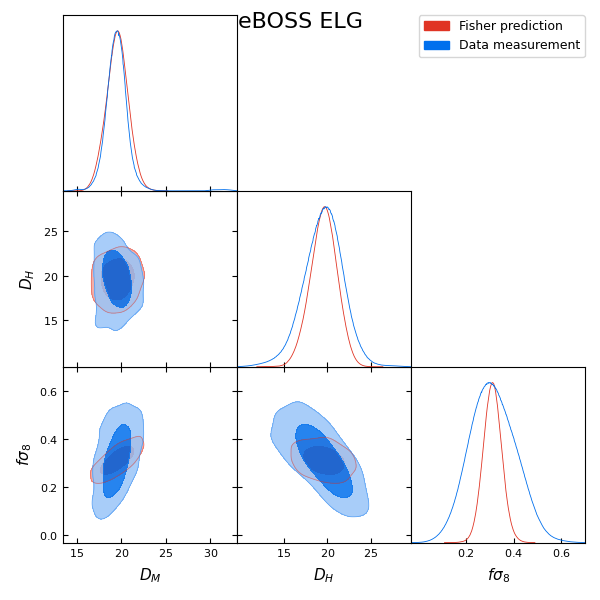}
    \caption{Constraints on $D_M$, $D_H$, $f\sigma_8$. Red contours show 68 and 95 percent confidence regions for the Fisher analysis with $k_{\rm max} = 0.09 \hompc$ (this work), and the blue contours show confidence regions for eBOSS ELG sample \cite{de_mattia_completed_2020}. Unlike other surveys, data and Fisher do not match very well in eBOSS ELG contour, as for this survey, we found that $k_{\rm max}^{\rm match} = 0.06 \hompc$, which is far from the average value: $k_{\rm max} = 0.09 \hompc$. For data, the grid of the relative probability has been used instead of the covariance matrix, as the ELG likelihood is not well-approximated as a Gaussian distribution.}
    \label{fig:eBOSS_ELG}
\end{figure}

We show contour plots for the data and Fisher matrices (with $k_{\rm max} = 0.09 \hompc$) for few
representative samples, spanning a range in redshift
and omitting some samples for clarity.
We show
BOSS DR12 Near (left panel Figure~\ref{fig:BOSS_Near_eBOSS_LRG}), eBOSS+CMASS LRG DR16 (right panel Figure~\ref{fig:BOSS_Near_eBOSS_LRG}), eBOSS Quasar DR16 (left panel Figure~\ref{fig:eBOSS_Quasar_DR16_WiggleZ}), WiggleZ Mid (right panel Figure~\ref{fig:eBOSS_Quasar_DR16_WiggleZ}), 
and, eBOSS Quasar DR14 (Figure~\ref{fig:eBOSS_Quasar_DR14}), and eBOSS ELG (Figure~\ref{fig:eBOSS_ELG}). The centres of these contours are set to the fiducial value for each parameter. If $b \sigma_8$ was not provided in the data covariance matrix, we omitted the $b\sigma_8$ contour for that survey.

\begin{center}

\begin{table}[h]
\centering
\caption{\label{tab:results3} The fractional error of $f \sigma_8^{\rm mg.\alpha s}$ in percent and the $k_{\rm max}$ at which observed (O.) and Fisher (F.) errors match. In contrast to Table \ref{tab:results}, the Fisher prediction errors are calculated after marginalizing over the dilation parameters (Mg.).}
\begin{tabular}{||c||c|c|cc|ccc|ccc||}
\hline\hline
  & 6dFGS
  & MGS 
  & \multicolumn{2}{c|}{BOSS} 
  & \multicolumn{3}{c|}{eBOSS} & \multicolumn{3}{c||}{WiggleZ}\\
  &
  & 
  & Near  &  Mid
  & $\rm LRG^{+}$ & ELG & QSO 
  & Near&	Mid&	Far
  \\
\hline

$f \sigma_8$(O.Mg.)&
-&  40.5&	7.8&	7.6&	7.9&
23.6&	9.3&	19.4&	16.1&	16.4	\\

$f \sigma_8$(F.Mg.)&

 -&   30.1&	7.8&	6.7&	6.5&	7.9&	7.3&	11.5&	9.5&	11.4	\\

\hline

$k_{\rm max}$(Match)&

-&   0.084&	0.122&	0.117&	0.113&	0.060&	0.104&	0.083&	0.086&	0.094\\

\hline\hline
 
\end{tabular}

\end{table}
\end{center}
\begin{center}

\begin{table}[h]
\centering

\caption{\label{tab:results4} Same as Table \ref{tab:results3} but including intermediate data releases.}
\begin{tabular}{||c||c|cc|cc|cc|cc|cc||}
\hline\hline
  & BOSS & \multicolumn{2}{l|}{BOSS DR10} & \multicolumn{2}{c|}{BOSS DR11} &
  \multicolumn{2}{c|}{BOSS DR12} &
  \multicolumn{2}{c|}{eBOSS DR14} &
  \multicolumn{2}{c||}{eBOSS DR16}
  \\

  & DR9
  & LZ & CM 
  & LZ & CM 
  & LZ & CM 
  & $\rm LRG^{+}$  & QSO
  & $\rm LRG^{+}$  & QSO
  \\
  
\hline

$f \sigma_8$(O.Mg.)&

14.6&   23.3&   12.8&   20.8&	9.9&   15.7&   8.6&	29.2&	16.4& 7.9& 9.3\\

$f \sigma_8$(F.Mg.)&

9.2&	12.9&	7.0&	11.1&	5.9&	10.6&	5.7&	10.5&	10.9&   6.5&    7.3\\

\hline

$k_{\rm max}$(Match)&
0.087&	0.073&	0.081&	0.072&	0.085&	0.082&	0.090&	0.054&	0.093& 0.113&	0.104 \\

\hline\hline
\end{tabular}
\end{table}
\end{center}

\clearpage
\printbibliography
\end{document}